\documentclass[authoryear,11pt]{article}

\usepackage{amsmath,supertabular,tabularx,graphicx,amssymb,amsfonts}
\usepackage{natbib,rotating}

\newcommand{\comments}[1]{}
\newcommand{\pcite}[1]{\citeauthor{#1}'s \citeyearpar{#1}}

\newtheorem{lemma}{Lemma}

\newtheorem{remark}{Remark}

\newcommand{\mby}{\textbf{y}}

\def\baro{\vskip  .2truecm\hfill \hrule height.5pt \vskip  .2truecm}
\def\barba{\vskip -.1truecm\hfill \hrule height.5pt \vskip .4truecm}

\title{A novel sandwich algorithm for empirical Bayes analysis of rank data}
\date{}
\author{Arnab Kumar Laha$^{1}$, Somak Dutta$^{2}$ and Vivekananda Roy$^{2}$\\
  $^1$ Indian Institute of Management, Ahmedabad, India,\\ $^2$ Department of Statistics, Iowa State
  University, USA
}

\begin{document}
\maketitle
\begin{abstract}
  Rank data arises frequently in marketing, finance, organizational
  behavior, and psychology. Most analysis of rank data reported in the
  literature assumes the presence of one or more variables (sometimes
  latent) based on whose values the items are ranked. In this paper we
  analyze rank data using a purely probabilistic model where the
  observed ranks are assumed to be perturbed versions of the true rank
  and each perturbation has a specific probability of occurring. We
  consider the general case when covariate information is present and
  has an impact on the rankings. An empirical Bayes approach is taken
  for estimating the model parameters. The Gibbs sampler is shown to
  converge very slowly to the target posterior distribution and we
  show that some of the widely used empirical convergence diagnostic
  tools may fail to detect this lack of convergence. We propose a
  novel, fast mixing sandwich algorithm for exploring the posterior
  distribution. An EM algorithm based on Markov chain Monte Carlo
  (MCMC) sampling is developed for estimating prior hyperparameters. A
  real life rank data set is analyzed using the methods developed in
  the paper. The results obtained indicate the usefulness of these
  methods in analyzing rank data with covariate information.
\end{abstract}


\section{Introduction}
The phenomenon of ranking a given set of items according to some
attributes is well known. Students are ranked routinely according to
their proficiency in some subject, job applicants are ranked according
to their suitability for a particular job, various brands of soap are
ranked by consumers according to their intention to purchase them at
some later time, various mutual funds are ranked by the investors
according to their intention to invest money in them etc. In some
cases, like ranking of students for their proficiency in a given
subject one may use some objective criterion like the marks scored in
an appropriately designed examination for assigning the rank - the student
who gets the highest marks gets rank 1, the student who gets the next
highest marks gets rank 2 etc. However, in other cases, like the case
of evaluating a job applicant, usually more than one attribute is
considered. Typically a panel of experts evaluate the items (the
candidates) on a variety of relevant dimensions some of which may be
objective while others are subjective. Often the individual experts are
only able to provide a ranking of the items as perceived by them
during the evaluation process.  The whole group of experts then
considers all the rankings by the individual experts and tries to
arrive at a consensus ranking through some subjective decision making
process. We call this generally accepted rank the ``true rank''. The
number of items to be ranked by an expert (a.k.a. subject) is
typically small since the ``robustness of ranking may well fall if the
subject is required to rank a large number of items simultaneously''
\cite[][p.19]{brow:1974}. Along the same lines \citet[][p. 81]{russ:gray:1994},
states that forcing subjects to give a different rank to each item may
``overtax their true powers of differentiation, with the result that
the rank differences they give do not reflect true value differences,
but merely add to the noise in the data.'' \citet{laha} provide a
statistical method to derive the ``true rank'' based on the rankings
provided by the individual experts. In this paper we extend this
method to the case when the true rank depends on some covariates.

When there are no covariates on the experts, we denote the true rank
as $\pi$ and the ranking given by the $i$th expert as $y_i$. The
ranking $y_i$ can be considered as a `perturbed' version of $\pi$ i.e.
we assume that $y_i=\sigma_i \circ \pi$ where $\sigma_i \in
\mathfrak{S}_p$, the set (group) of permutations of the integers
$1, 2, \dots, p$.  The permutations $\sigma_i$'s are assumed to be
independently distributed $\mathfrak{S}_p$ valued random variables
following some common law.  \citet{laha} discuss the estimation of
$\pi$ using a Bayesian approach and the Sampling Importance Resampling
(SIR) technique.

Suppose now that there are covariates on the experts and the true rank
depends on the value of the covariate. Let $X$ be the covariate and
suppose it takes the value $x_i$ for the $i$th expert. In this case, we
have the model $y_i=\sigma_i \circ \pi(x_i)$ where as before
$\sigma_i$'s are i.i.d. $\mathfrak{S}_p$ valued random variables. We are
interested in estimating $\pi(x)$ where $x \in Range(X)$.

There are several other methods in the literature for analyzing rank
data most of which do not incorporate covariates. Models based on
order-statistics are discussed in \citet{thurstone}, \citet{yellot},
\citet{critchlow80}, \citet{daniels}, \citet{mosteller} and
\citet{yu}. Order-statistics based models consider a latent utility
$y_i$, of $i$th item, $i=1,\ldots,p$ given by a expert who ranks the
items in decreasing order of utility. While some of them assume that
for a given expert the variables $y_i$'s are independent, \citet{yu}
considers a multivariate normal model for the utilities with
non-identity covariance matrix. He also considers the case when
covariates are associated with items and experts. Models based on
distance are first considered in \citet{mallows}. These models are
further studied in \citet{feigincohen} and \citet{diaconis}. Further
work in distance based models has been reported in
\citet{fligner86}. Mallows model has been generalized by
\citet{lebanon:cranking}. \citet{fligner88} proposed multistage
ranking model where they introduced the concept of \emph{central
  ranking} $\pi$ but without any covariates. Recently, mixtures of
distance-based models have been considered in \citet{murphy}. An
approach based on mixture of experts model to cluster voters in Irish
election background has been recently considered in
\citet{gormley}. The mixture of experts model can include covariates
\cite[see also][]{gormley10}. Also other methods for analyzing rank
data without covariates using group representation have been studied
in \cite{diaconis89}. However, our approach of estimating the
\emph{true rank} is not considered elsewhere and is innovative in this
regard.

We develop a flexible conjugate prior for Bayesian inference. The use
of conjugate prior allows us to construct a Gibbs sampler with
standard conditional distributions. Although it is shown that the
Gibbs sampler is a {\it uniformly ergodic} Markov chain, it converges
very slowly to the target posterior distribution because it moves
between the modes of the posterior too infrequently \cite[see e.g.][
for definition of uniformly ergodic Markov chain]{meyn:twee:1993}.  On
the other hand, this lack of convergence is not detected by a popular
Markov chain Monte Carlo (MCMC) empirical convergence diagnostic tool,
namely the potential scale reduction factor introduced by \cite{gelm:rubi:1992}. Even
though the Gibbs sampler is stuck in a local mode, the autocorrelation
(between iterations of the Gibbs chain) is close to zero even before
lag three. Only when the Gibbs sampler is run for a very large number
(more than five million) of iterations, it visits other modes and the
autocorrelations jump to near one showing lack of mixing.  This is an
alarming issue as in practice, MCMC users employ these empirical
convergence diagnostic tools to determine MCMC simulation length and
we show that inference drawn from prematurely stopped MCMC simulation
may be far from the truth. These observations demonstrate the danger
in depending purely on empirical convergence diagnostic tools to
verify convergence of MCMC algorithms as these tools may give false
indication of convergence.

Every two-variables Gibbs sampler has the same {\it rate of
  convergence} as its two reversible subchains \cite[see
e.g.][]{robe:case:2004, liu:wong:kong:1994}. These subchains can also be viewed as data
augmentation (DA) chains \cite[see][]{hobe:marc:2008}.
Over the last two decades a lot of effort has gone into modifying DA
chains to improve its convergence. These improved algorithms include
the parameter expanded DA (PX-DA) algorithm of \citet{liu:wu:1999},
the conditional and marginal augmentation algorithms of
\citet{meng:vand:1999}, the sandwich algorithm of
\citet{hobe:marc:2008} and finally the interweaving algorithms of
\citet{yu:meng:2011}. Here we consider \pcite{hobe:marc:2008} sandwich
techniques to improve the subchains of our Gibbs sampler. In a sandwich algorithm, an
extra step is added (sandwiched) to the Gibbs sampler in between the
draws from the conditional distributions. 
We construct a sandwich algorithm and use the
Rao-Blackwellization technique to make inference about the true rank,
$\pi$. Using results in \cite{hobe:roy:robe:2011} we show that all the
(ordered) eigenvalues of the Markov operator corresponding to the
sandwich chain are less than or equal to the corresponding eigenvalues
of the DA Markov operator. We also present an EM algorithm based on
MCMC sampling to make inference about the prior hyperparameter.

The structure of the paper is as follows: In
Section~\ref{sec:algprob}, we introduce the probabilistic rank model
with covariates and discuss estimation of the model parameters using a
Bayesian approach.  In Section~\ref{sec:modelspecification}, we build
prior distributions on the parameters and provide the joint and
marginal posterior distributions (up to normalizing
constants). Section~\ref{sec:comp} contains construction of MCMC
algorithms as well as the EM algorithm for estimating the prior
hyperparameter. Section~\ref{sec:simu} demonstrates the convergence
issues of the Gibbs sampler through simulation examples. We also describe
how our sandwich algorithm results in huge improvement in mixing by
adding an extra step to facilitate moves between the modes of the posterior.
In Section~\ref{sec:data}, we discuss analysis of a real life
dataset using methods developed earlier in the paper. Some concluding
remarks are given in Section~\ref{sec:disc}.

\section{An algebraic--probabilistic model}\label{sec:algprob}
Let us think of a situation in which $p$ items are ranked by a random
sample of $n$ experts from a population. In the absence of any
covariate information, \citet{laha} suppose that there
is a ``true rank" $\pi$ of the $p$ items and the observed ranks are
``perturbed" versions of the true rank $\pi$. The perturbation is the
composition of a random permutation from the group of all permutations
$\mathfrak{S}_{p}$ with the true rank $\pi.$ When there is one or more
categorical variable partitioning the population of the experts into
few categories, a natural way to extend the model is to assume that
there is a true rank associated with each category of experts and
the observed ranks are some random perturbed versions of these true
ranks. Thus, if there are $g$ categories of experts and $b_j$
denotes the number of experts in the $j$th category and $y_{ij}$'s
are the ranks observed from the experts in $j$th category, then we
assume that there are true ranks $\pi_1,\pi_2,\ldots,\pi_g $ such that
\begin{equation}\label{model:APmodel}
y_{ij} = \sigma_{ij} \circ \pi_j,\quad i=1,\ldots,b_j,
\end{equation}
for $j=1,\ldots,g,$ where $\sigma_{ij}$'s are i.i.d random
permutations having some distribution and are synonymous to random
error perturbations. Notice that because composition is the natural
group operation on the group $\mathfrak{S}_p$, \eqref{model:APmodel}
is the natural ANOVA model on the permutation group.

We use the following notations to denote the $p!$ permutations in
$\mathfrak{S}_p$.
Let $\zeta_j = j$th ranking in lexicographic order in
$\mathfrak{S}_p$. So $\zeta_1$ is the identity permutation,
\[
\zeta_2 = \left( \begin{array}{ccccc} 1 & 2 & \cdots & p-1 & p \\ 1 &
    2 & \cdots & p & p-1\end{array} \right),\]
\[  \zeta_3 =
\left( \begin{array}{cccccc} 1 & 2 & \cdots & p-2 & p-1 & p \\ 1 & 2 &
    \cdots & p-1 & p-2 & p\end{array} \right)\] and so on. Finally, \[
\zeta_{p!} = \left( \begin{array}{ccccc} 1 & 2 & \cdots & p-1 & p \\ p
    & p-1 & \cdots & 2 & 1\end{array} \right) .\]

Next, in order to construct the distribution of the $\sigma_{ij}$'s,
suppose they take the value $\zeta_k$ with probability $\theta_k,$
i.e. $P(\sigma = \zeta_k) = \theta_k.$ It is expected that the
experts are sensible so that \emph{large} random error
perturbations are less likely to occur than the \emph{smaller} ones,
where we order the permutations by their distance from the identity
permutation. Thus, $\theta_1$ should be the highest because it is the
probability that the observed rank equals to the true rank for any
category. Moreover, whenever $\zeta_k$ is farther away from $\zeta_1$
than $\zeta_l,$ it is expected that $\theta_k$ would be smaller than
$\theta_l.$ Based on this assumption, we build our prior distribution
on the vector $\boldsymbol\theta = (\theta_1,\ldots,\theta_{p!})$ in
the next section.

We break our task into three parts. First we construct prior
distributions on the parameters necessary for the Bayesian
inference. Next we derive the analytic forms of the marginal posterior
distributions of the true ranks $\pi_1,\ldots,\pi_g$ and the vector
$\boldsymbol\theta,$ up to a normalizing constant and demonstrate
acute challenges present in computing the normalizing constant. In
order to avoid the mammoth computational challenge, we construct a
powerful MCMC algorithm that enables us to make fast inference.

\subsection{Prior distributions}
\label{sec:modelspecification}
In order to construct a prior on $\boldsymbol\theta$ reflecting our
belief that the experts are less likely to make big errors, we
begin with fixing the notion of big or small perturbations using the
Cayley distance on $\mathfrak{S}_{p}$ \citep{cayl:1849}. The Cayley distance between two
permutations $\tau$ and $\alpha$ is defined as
\[d_C(\tau,\alpha) = p - |\tau\circ\alpha^{-1}|\] where
$|\tau\circ\alpha^{-1}|$ is the number of cycles in the unique
representation of $\tau\circ\alpha^{-1}$ as a product of disjoint
cycles.  Based on this distance, we call a perturbation $\tau$ smaller
than a perturbation $\alpha,$ if $d_C(\tau,\zeta_1) <
d_C(\alpha,\zeta_1).$ Consequently, we take the prior on $\boldsymbol\theta$ as a
Dirichlet$(a_1,\ldots,a_{p!})$ distribution because it is the
conjugate prior. We set the hyperparameters of this distribution as
\begin{equation} \label{eqn:thetaHyperParam}
a_k = \exp(\lambda~(p-d_C(\zeta_k,\zeta_1))) = \exp(\lambda|\zeta_k|),
\end{equation}
where $\lambda>0$ is a parameter that reflects the overall precision
in ranking the items. If $\lambda$ is large then the distribution of
the error $\sigma$ is concentrated on the set of permutations having
many cycles and thus at most a few items are mis-ranked by the
experts. In contrast, when the value of $\lambda$ is small, the
experts end up mis-ranking a large number of items. This prior
distribution on the $\boldsymbol\theta$ stems from the natural
exponential family on the permutation group \citet{mccullagh2011}
which puts a mass proportional to $\exp(\lambda|\zeta_k|)$ on the
permutation $\zeta_k.$ However, had we deterministically set
$\theta_k \varpropto \exp(\lambda|\zeta_k|)$, computations would have
been very difficult because the nice conjugacy structure enjoyed by
our model would be broken.

In this article, we follow an empirical Bayesian route and estimate
the hyperparameter $\lambda$ by maximizing the marginal likelihood
$p(\mby | \lambda),$ the details of which are given in Section
\ref{sec:estimateLambda}. Estimating $\lambda$ by maximizing the marginal likelihood
ensures that the inference remains invariant to multiplying the
$a_k$'s in \eqref{eqn:thetaHyperParam} by any arbitrary
constant. 

As far as the prior on the true ranks are concerned, we could incorporate
any available knowledge while constructing the prior distribution. For
example, if an item is known to be favorite we can construct a prior
under which that particular item is more likely to be a top choice. If
there is any reason to believe spatial correlation among the true
ranks, e.g. for voting data, we can incorporate this information in
the prior. For simplicity and ease of computation, however, we would
like to assume that the $\pi_j$'s are apriori independently
distributed and if there is no obvious reason as to why one or more
items should be given preference to others, we can keep uniform priors
(on $\mathfrak{S}_{p}$). In general, we denote the prior on
$\boldsymbol\pi \equiv (\pi_1,\pi_2,\ldots,\pi_g)$ by $p(\boldsymbol\pi)=
p_1(\pi_1)p_2(\pi_2)\cdots p_g(\pi_g).$

\begin{figure*}[htp]
\centering\includegraphics[width=0.6\textwidth]{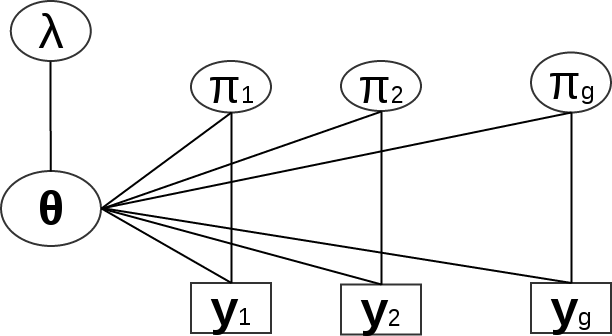}
\caption{Graphical representation of the algebraic--probabilistic
  model proposed in this paper. Here $\mathbf{y}_j = \{y_{ij},i=1,\ldots,b_j\},$ is the
  collection of all ranks given by experts in the $j$th category;
  $j=1\ldots,g.$}
\label{fig:graphical}
\end{figure*}

\subsection{Exact posterior and its limitations}\label{sec:exactPosterior}
It is quite useful to visualize the conditional dependence structure
among the variables. This is displayed in Figure
\ref{fig:graphical}. Two features are noticeable from the
graph. Firstly, $\pi_j$'s are conditionally independent
among themselves given $\boldsymbol\theta$ and the data $\mby.$
Secondly, the likelihood of $\lambda$ given all other variables
depends only on $\boldsymbol\theta.$ These two observations will be
key ingredients in our computational methods later in the article. To
this end, notice that the likelihood of the data is given by
\begin{equation}\label{eqn:lik}
p(\mby |~\boldsymbol\theta,\pi_1,\ldots,\pi_g) = \prod_{j=1}^{g}\prod_{i=1}^{p!}\theta_{\zeta_i\circ \pi_j^{-1}}^{n_{ij}}
\end{equation}
where $n_{ij}$ is the number of times the ranking $\zeta_i$ is
observed in the $j$th category. Since the conjugate Dirichlet prior for
$\boldsymbol\theta$ is assumed, the joint posterior simplifies to
\begin{equation}\label{eqn:posterior}
  p(\boldsymbol\theta,\pi_1,\ldots,\pi_g | \mby,\lambda) \varpropto \prod_{k=1}^{p!} \theta_{k}^{m_k(\boldsymbol\pi) + a_k - 1}  \times p(\boldsymbol\pi),
\end{equation}
where \[m_k(\boldsymbol\pi) ~=~ \sum_{j=1}^g
~\sum_{i=1}^{p!}~n_{ij}~ \mathbb{I} (\zeta_i \circ \pi_j^{-1} =
\zeta_k),\] where $\mathbb{I}(\cdot)$ is the indicator
function. Integrating out $\boldsymbol\theta$ gives the marginal
posterior of $\boldsymbol\pi = (\pi_1,\ldots,\pi_g)$ up to a
normalizing constant:
\begin{equation}\label{eqn:posterior_pi}
  p(\boldsymbol\pi | \mby,\lambda) \varpropto p(\boldsymbol\pi) \prod_{k=1}^{p!}\Gamma(m_k(\boldsymbol\pi) + a_k).
\end{equation}
 Further, the marginal (joint) posterior of $\boldsymbol\theta$
is given by a mixture of Dirichlet densities,
\begin{equation}\label{eqn:posterior_theta}
 p(\boldsymbol\theta | \mby,\lambda) = \sum_{\boldsymbol\pi \in \mathfrak{S}_p^g} \mathcal{D}(\boldsymbol\theta;m_1(\boldsymbol\pi)+a_1,\ldots,m_{p!}(\boldsymbol\pi) + a_{p!}) p(\boldsymbol\pi | \mby,\lambda),
\end{equation}
where $\mathfrak{S}_p^g = \mathfrak{S}_p\times \mathfrak{S}_p \times \cdots \times \mathfrak{S}_p,$ $g$ times.
From \eqref{eqn:posterior_theta} the marginals of $\theta_k$, for $k=1,...,p!$ can be
obtained as mixtures of Beta densities,
\begin{align*}\label{eqn:posterior_theta_individual}
  p(\theta_k | \mby,\lambda) &= \sum_{\boldsymbol\pi\in \mathfrak{S}_p^g} \mathcal{B}(\theta_k; m_k(\boldsymbol\pi)+a_k,N +a_0 - m_k(\boldsymbol\pi) - a_k)\\ & \times p(\boldsymbol\pi | \mby,\lambda),
\end{align*}
where $N$ is the total sample size and $a_0 = \sum_{k\geq 1}
a_k$. One important thing stands out from these exact posterior
distributions. The mixture representations of the marginal posterior
densities of $\theta_k$'s indicate that they might be
multimodal. However, unless the rankings are completely random, only a
few of the components in the mixture have substantial
contributions. That is, $p(\boldsymbol\pi | \mby,\lambda)$'s are
insignificant for all but few $\boldsymbol\pi = (\pi_1,\ldots,\pi_g)
\in \mathfrak{S}_p^g.$ Thus a subjective judgment about goodness of
fit can be deduced from the number of modes of posterior marginals of
$\theta_k$'s. If there are too many modes, this would indicate that
the model is not good for that data.

However, these exact forms are not of much use in real applications
mainly because the normalizing constant in \eqref{eqn:posterior_pi}
requires too much computational effort. Even when $p,$ the number of
items is small, the storage requirement and computational complexity
are $O(p!^g)$ which scale exponentially in $g$. For example, if there
are few factorial variables present and each of them have a moderate
number of levels, the total number of categories, i.e. $g$, will be
large. Thus when there are only $p = 3$ items to be ranked and there
are only $g = 12$ categories, a total of 16GB of memory is required to
store the values in \eqref{eqn:posterior_pi}, that are required to not
only compute the normalizing constant but also the marginal posteriors
of $\theta_i$'s. With $p=4,$ this computational bottleneck is reached
even for $g$ as low as 7. Note that for the sushi data example considered
in Section~\ref{sec:data} the memory requirements would have been $10^{24}$ GB.

Yet, despite their futility in direct computation, these formulas pave
way for constructing a Gibbs sampler because the full conditionals
consist of tractable distributions due to conjugacy that is displayed
in \eqref{eqn:posterior}. The main reasons behind constructing a Gibbs
sampler is that the per iteration computational and storage
requirements scale linearly in both $p!$ and $g.$ Thus even if $g$ is
large, a Gibbs sampler remains a practically viable option as long as
it converges fast. However, as we shall see the traditional Gibbs
sampler has a very slow convergence rate. And thus we shall construct
a novel and efficient sandwich algorithm by using \eqref{eqn:posterior_pi}.

\section{MCMC algorithms and parameter estimation}
\label{sec:comp}
In this section we construct MCMC algorithms for making inference on
the model parameters. The conditional densities and calculations in
sections~\ref{sec:mcmc} and \ref{sec:imda} are conditional on
$\lambda$. Then in section~\ref{sec:estimateLambda} we describe an EM
algorithm for estimating the hyperparameter $\lambda$.

\subsection{The Gibbs sampler}
\label{sec:mcmc}
We begin with deriving the conditional distributions associated with
the joint posterior density \eqref{eqn:posterior} in order to
construct a Gibbs sampler. Notice that the conditional distribution
of $\boldsymbol \theta$ given $\boldsymbol \pi$ and $\mby$ is given by
\[
p( \boldsymbol \theta | \boldsymbol \pi, \mby) \propto
\prod_{k=1}^{p!} \theta_{k}^{m_k(\boldsymbol\pi) + a_k - 1},
\]
that is, $\boldsymbol \theta | \boldsymbol \pi, \mby$ has a Dirichlet
distribution.  If independent priors are used on $\pi_1,\ldots,\pi_g$,
from \eqref{eqn:posterior} we see that the conditional distribution of $\boldsymbol \pi$
given $\boldsymbol \theta$ and $\mby$ is
\begin{align*}
p(\pi_1,&\ldots,\pi_g | \boldsymbol\theta, \mby) \propto \prod_{k=1}^{p!} \theta_{k}^{m_k(\boldsymbol\pi) + a_k - 1} p_1(\pi_1)\cdots p_g(\pi_g)\\
&\propto \prod_{j=1}^g\prod_{k=1}^{p!}\theta_k^{\sum_i n_{ij} \mathbb{I} (\zeta_i \circ \pi_j^{-1} = \zeta_k)}\times p_1(\pi_1)\cdots p_g(\pi_g),
\end{align*}
that is, $\pi_1,\ldots,\pi_g$ are conditionally independent. Thus
\begin{equation}
  \label{eq:condpigith}
  p(\pi_j = \zeta_r | \boldsymbol\theta, \mby) \propto​\prod_{k=1}^{p!} \theta_k^{\sum_i n_{ij}\mathbb{I} (\zeta_i  = \zeta_k\circ\zeta_r)}\times p_j(\zeta_r) \equiv \gamma_{jr} (\boldsymbol \theta),
\end{equation}
 $r=1,\ldots,p!$. So conditional on $\boldsymbol \theta$ and $\mby$, $\pi_1,\ldots,\pi_g$ are independent
multinomial random variables. Note that, for fixed $k$ and $r$, there is only one
$i$ satisfying $\zeta_i  = \zeta_k\circ\zeta_r$, and it can be found outside
the Gibbs sampler simulation as it does not change between iterations.
Let $(\boldsymbol\theta^{(m)}, \boldsymbol\pi^{(m)})$ be the current value
of the Gibbs sampler at step $m$, then the following two steps are
used to move to $(\boldsymbol\theta^{(m+1)}, \boldsymbol\pi^{(m+1)})$:

\baro \vspace*{2mm}
\noindent {\rm Iteration $m+1$ of the Gibbs sampler:}
\begin{enumerate}
\item Draw $ \boldsymbol \pi^{(m+1)} \sim p(\cdot|\boldsymbol\theta^{(m)},
  \mby)$ which can be done as follows. For $j=1,2,\dots,g$, normalize
  $\gamma_{j1} (\boldsymbol\theta^{(m)}),\dots, \gamma_{jp!}
  (\boldsymbol\theta^{(m)})$ so that they add up to 1, then draw $\pi_j^{(m+1)} \sim$ multinomial $(1;
  \gamma_{j1}^{(m)}(\boldsymbol\theta^{(m)}), \dots, \gamma_{jp!}^{(m)}(\boldsymbol\theta^{(m)}))$.
\item Draw $ \boldsymbol\theta^{(m+1)}| \boldsymbol \pi^{(m+1)},
  \mby \sim $ Dirichlet $(m_1(\boldsymbol\pi^{(m+1)}) + a_1, \dots,m_{p!}(\boldsymbol\pi^{(m+1)}) + a_{p!})$.
  \vspace*{-3.5mm}
\end{enumerate}
\barba \bigskip

\noindent For $k \ge 1$, define
\[
S_k := \Big\{ \boldsymbol \theta \in \mathbb{R}^k: \theta_i \in [0,1] \; \; \mbox{and} \;\;
\theta_1+\cdots+\theta_k=1 \Big\} \;.
\]
\noindent The above mentioned Gibbs algorithm results in a Markov
chain $\{\boldsymbol \theta^{(m)}, \boldsymbol \pi^{(m)}\}_{m \ge 0}$
with state space $S_{p!}  \times \mathfrak{S}_p^g$ and invariant
density $p(\boldsymbol \theta, \boldsymbol \pi | \mby, \lambda)$ given
in \eqref{eqn:posterior}. It is known that the (sub) chains
$\{\boldsymbol \pi^{(m)}\}_{m \ge 0}$ and
$\{\boldsymbol\theta^{(m)}\}_{m \ge 0}$ are reversible Markov chains
\cite[][Lemma 9.11]{robe:case:2004}. Since the conditional densities
$p(\boldsymbol\pi | \boldsymbol\theta, \mby)$ and $p(\boldsymbol\theta
| \boldsymbol\pi, \mby)$ are everywhere positive, it implies that the
Markov chain $\{\boldsymbol\pi^{(m)}\}_{m \ge
  0}$ is irreducible and aperiodic. Because
$\{\boldsymbol\pi^{(m)}\}_{m \ge 0}$ is a finite state space Markov
chain, it follows that $\{\boldsymbol\pi^{(m)}\}_{m \ge 0}$ is
uniformly ergodic with unique invariant density $p(\boldsymbol\pi |
\mby, \lambda)$ given in \eqref{eqn:posterior_pi}.  Moreover, it is well known that
$\{\boldsymbol\theta^{(m)}, \boldsymbol\pi^{(m)}\}_{m \ge 0}$ as well
as its two subchains $\{\boldsymbol\pi^{(m)}\}_{m \ge 0}$ and $\{\boldsymbol\theta^{(m)}\}_{m \ge 0}$ converge to
their respective invariant distributions at the same rate \cite[see
e.g.][]{liu:wong:kong:1994}. Hence we have the following lemma.
\begin{lemma}
The Gibbs chain $\{\boldsymbol\theta^{(m)}, \boldsymbol\pi^{(m)}\}_{m \ge
0}$ and the marginal chains $\{\boldsymbol\pi^{(m)}\}_{m \ge
0}$, $\{\boldsymbol\theta^{(m)}\}_{m \ge
0}$ are uniformly ergodic.
\end{lemma}

In Section~\ref{sec:simu} we consider the performance of the sub
chains and hence the Gibbs sampler through some simulation examples
and observe that the Gibbs sampler suffers from slow convergence due
to its inability to move between local modes. In the next section, we
construct an algorithm improving the Markov chain
$\{\boldsymbol\theta^{(m)}\}_{m \ge 0}$.

\subsection{Improving the DA chain $\{\boldsymbol\theta^{(m)}\}_{m \ge
0}$}
\label{sec:imda}
The $\boldsymbol \theta$ subchain, $\{\boldsymbol\theta^{(m)}\}_{m \ge
  0}$, of the Gibbs sampler is a Markov chain with stationary density
$p(\boldsymbol\theta | \mby, \lambda)$ given in
\eqref{eqn:posterior_theta}. This chain can be viewed as a DA chain.
As mentioned in the introduction, here we consider
\pcite{hobe:marc:2008} sandwich technique to improve this DA chain. A
sandwich algorithm (SA) is a simple alternative to the DA algorithm
that often converges much faster.  Each iteration of a generic SA has
three steps.  Let $r(\boldsymbol \pi'| \boldsymbol \pi)$ be a Markov transition density (Mtd)
with invariant density $p(\boldsymbol \pi | \mby, \lambda)$ given in
\eqref{eqn:posterior_pi}.  If $\tilde{\boldsymbol\theta}^{(m)}$ is the current
value of the sandwich chain at step $m$, then three steps are used to
move to the new state $\tilde{\boldsymbol\theta}^{(m+1)}$--- draw $
\boldsymbol\pi \sim p(\boldsymbol\pi | \tilde{\boldsymbol\theta}^{(m)},
\mby)$, draw $\boldsymbol\pi' \sim r(\cdot|\boldsymbol\pi)$, and
finally draw $ \tilde{\boldsymbol\theta}^{(m+1)} \sim p(\boldsymbol\theta
|\boldsymbol\pi', \mby)$.  A routine calculation shows that the
sandwich chain remains invariant with respect to $p(\boldsymbol\theta
| \mby, \lambda)$, so it is a
viable alternative to the DA chain. Note that the first and the last
steps of the SA are exactly the two steps used in the Gibbs sampler
presented in section~\ref{sec:mcmc}. \cite{roy:2012a} shows that the
sandwich chains {\it always} converge at least as fast as the
corresponding DA algorithms. In particular, \cite{roy:2012a} shows
that SA is at least as good as the DA in terms of having smaller
operator norm \cite[see also][]{hobe:marc:2008}. Clearly, on a per
iteration basis, it is more expensive to simulate the sandwich chain
than the DA chain. However, it may be possible to find an $r$ that
leads to a huge improvement in mixing despite the fact that the
computational cost of drawing from $r$ is negligible relative to the
cost of drawing from $p(\boldsymbol\theta|\boldsymbol\pi, \mby)$ and
$p(\boldsymbol\pi | \boldsymbol\theta, \mby)$ \cite[see
e.g.][]{roy:2014, roy:hobe:2007, hobe:roy:robe:2011}.

We propose the following SA improving the $\boldsymbol\theta$ subchain,
$\{\boldsymbol\theta^{(m)}\}_{m \ge 0}$, of the Gibbs sampler. Let
$\tilde{\boldsymbol\theta}^{(m)}$ be the current value of the sandwich
chain at step $m$, then the following three steps are used to move to
$\tilde{\boldsymbol\theta}^{(m+1)}$:

\baro \vspace*{2mm}
\noindent {\rm Iteration $m+1$ of the Sandwich Algorithm:}
\begin{enumerate}
\item Draw $ \boldsymbol\pi \sim p(\cdot|\tilde{\boldsymbol\theta}^{(m)}, \mby)$ as in step 1 of the Gibbs sampler.
\item Draw a permutation $\sigma$ randomly uniformly from $\mathfrak{S}_p$. Move to
$\boldsymbol\pi' = (\sigma \circ \pi_1, \sigma \circ \pi_2, \dots, \sigma \circ \pi_g) $ with probability
\[
\mbox{min}\Big\{1, \frac{p(\boldsymbol\pi' | \mby, \lambda)}{p(\boldsymbol\pi, | \mby, \lambda)}\Big\},
\]
otherwise $\boldsymbol\pi$ is retained.
\item Draw $\tilde{\boldsymbol\theta}^{(m+1)} \sim p(\cdot| \boldsymbol\pi' , \mby)$ as in step 2 of the Gibbs sampler.
\vspace*{-3.5mm}
\end{enumerate}
\barba \bigskip
\noindent Although the proposed SA is computationally slightly more demanding
than the Gibbs sampler, it shows huge gain in mixing in both the
simulation examples presented in Section~\ref{sec:simu} as well as the
real data examples in Section~\ref{sec:data}. In
Section~\ref{sec:simu}, we see that the extra sandwich step ($r$)
helps the chain move between local modes. The above idea of using random permutation for facilitating moves between
local modes is similar to the permutation sampler \citep{fruh:2001} developed for
Bayesian mixture models although there are differences. Firstly in \cite{fruh:2001} the
random label switching step is applied on the latent variables, and thus as explained in
\cite{hobe:roy:robe:2011} it fits directly into the SA setup of \cite{hobe:marc:2008}. Whereas here the
random permutation is applied on the parameters $\boldsymbol\pi$ and this is why we use a Rao
Blackwellized estimator of $\boldsymbol\pi$ based on the sandwich chain $\{ \tilde{\boldsymbol\theta}^{(m)}\}$.
Secondly, the permutation sampler samples from the so-called unconstrained posterior in
the mixture model. Thus extra post processing (satisfying identifiability constraint) is needed
to make valid inference on the parameters---here we do not need such extra computation.

\begin{remark}
  \label{rem:reduc}
  Since the sandwich step, $r(\boldsymbol\pi'| \boldsymbol\pi)$, is a
  Metropolis Hastings (MH) step, it is reversible with respect to
  $p(\boldsymbol\pi | \mby, \lambda)$ and hence has $p(\boldsymbol\pi
  | \mby, \lambda)$ as its invariant density. On the other hand, the
  chain driven by $r$ is reducible and can move to at most $p!$ many
  states. (Recall that the cardinality of the state space of
  $\{\boldsymbol\pi^{(m)}\}_{m \ge 0}$ is $(p!)^g$.) The reducibility
  of $r$ is common among efficient sandwich chains \citep{roy:2012b, hobe:roy:robe:2011, roy:hobe:2007}.
\end{remark}

\begin{remark}
  \label{rem:mhalt}
  The step 2 of our SA may suggest an alternative algorithm for
  sampling from \eqref{eqn:posterior} where in each iteration an
  ({\it irreducible}) MH step is used for the marginal
  \eqref{eqn:posterior_pi} followed by a draw from the conditional
  density $p(\boldsymbol\theta |\boldsymbol\pi, \mby)$ as in step 3 of
  the SA. We tried to implement this algorithm with different MH
  proposals including the uniform distribution. But, the average
  acceptance probability was too low for these algorithms to be
  practical.
\end{remark}
 As mentioned before, any Mtd
$r$ with invariant distribution $p(\boldsymbol\pi | \mby, \lambda)$
can be used to construct a valid SA. When $p$ is large, then a
local move for $\pi'$ may be preferred for the sandwich step.  For
example, in this case we can generate $\tau$
from a distribution which gives high probability on small
permutations but zero probability on the identity permutation. (One
choice may be to use the multinomial distribution with parameter $a_k$
defined in Section~\ref{sec:modelspecification} with the restriction
$a_1= 0$.) We then propose permutations $\pi'_j = \tau \circ \pi_j$
for $j=1, 2, \dots, g$, which is accepted with the corresponding
MH acceptance probability.

As mentioned before, SA is always at least as good as the DA in terms
of having smaller operator norm. \cite{hobe:roy:robe:2011} mention
that although the norm of a Markov operator provides a univariate
summary of the convergence behavior of the corresponding chain, a
detailed picture of the convergence can be found by studying the
spectrum of the Markov operator. (See \cite{hobe:roy:robe:2011} for a
gentle introduction to Markov operators and their spectrum.)
Lemma~\ref{prop:sa} shows that the spectrum of our sandwich
chain dominates that of the $\boldsymbol\theta$ subchain of the Gibbs
sampler in the sense that all the (ordered) eigenvalues of the SA are
at most as large as the (ordered) eigenvalues of the later. Let
$L^2_0(p(\boldsymbol\theta | \mby, \lambda))$ denotes the vector space
of all mean zero, square integrable (with respect to
$p(\boldsymbol\theta | \mby, \lambda)$) functions defined on $S_{p!}$.
Let $K_{\boldsymbol\theta}$ and $\tilde{K}_{\boldsymbol\theta}$ be the
Markov operators, $L^2_0(p(\boldsymbol\theta | \mby,
\lambda)) \rightarrow L^2_0(p(\boldsymbol\theta | \mby,
\lambda))$, corresponding to $\{\boldsymbol\theta^{(m)}\}_{m \ge 0}$
and the sandwich chain, $\{\tilde{\boldsymbol\theta}^{(m)}\}_{m \ge
  0}$ respectively. Since the sandwich step $r$ is performed based on
a uniform draw from $\mathfrak{S}_p$, it follows that two consecutive
steps from $r$ still results in a uniform draw. Thus $r$ is idempotent
and the following lemma follows from Theorem 1 of \cite{hobe:roy:robe:2011}. Let $q \equiv (p!)^g$.
\begin{lemma}
  \label{prop:sa}
The operators $K_{\boldsymbol\theta}$ and $\tilde{K}_{\boldsymbol\theta}$ are
  both compact and each has a spectrum that consists exactly
  of the point $\{0\}$ and $q -1$ eigenvalues in $[0,1)$.  Furthermore,
  if we denote the eigenvalues of $K_{\boldsymbol\theta}$ by
\[
0 \le \rho_{q-1} \le \rho_{q-2} \le \cdots \le \rho_1 < 1 \;,
\]
and those of $\tilde{K}_{\boldsymbol\theta}$ by
\[
0 \le \tilde{\rho}_{q-1} \le \tilde{\rho}_{q-2} \le \cdots \le
\tilde{\rho}_1 < 1 \;,
\]
then $\tilde{\rho}_i \le \rho_i$ for each $i \in
\{1,2,\dots,q-1\}$.
\end{lemma}
Since the conditional probabilities $P(\pi_i = \zeta_j |
\boldsymbol\theta, \mby)$ are available in closed form (see
\eqref{eq:condpigith}), we use the Rao-Blackwellized estimator based
on the sandwich chain $\{\tilde{\boldsymbol\theta}^{(m)}\}_{m =
  0}^M$ for estimating the true rank probabilities, that
is,
\[
\hat{P}(\pi_i = \zeta_j |\mby) = \frac{1}{M}\sum_{m=1}^M P(\pi_i = \zeta_j | \tilde{\boldsymbol\theta}^{(m)}, \mby).
\]
Given a sample of $\boldsymbol \theta$'s (using the sandwich chain)
from its marginal posterior density we can get a sample from the
posterior of $\boldsymbol \pi$, by drawing from the conditional
distributions of $\pi_j$'s given $\boldsymbol \theta, j=1,\dots,g$. We can use this
sample to estimate marginal as well as joint probability distributions
of $\pi_j$'s.
\begin{remark}
  \label{rem:sapi}
  In this section, an SA is constructed improving the $\boldsymbol\theta$ sub
  chain of the Gibbs sampler. Unfortunately, we could not construct
  (except for small $p$ and $g$) a sandwich chain that converges
  faster than the $\boldsymbol\pi$ subchain $\{\boldsymbol\pi^{(m)}\}_{m \ge 0}$
  of the Gibbs sampler and at the same time the computational cost for
  simulating it is similar to that for $\{\boldsymbol\pi^{(m)}\}_{m
    \ge 0}$. The difficulty of constructing such an SA is due
  to the intractability of the marginal posterior density
  $p(\boldsymbol\theta | \mby, \lambda)$.
\end{remark}
We now briefly discuss some of the useful
properties of our proposed model that are utilized in the data analysis in Section~\ref{sec:data}.

\subsection{Computing joint and conditional posterior probabilities}
\label{sec:compjtcond}
A novelty of the model and the sandwich algorithm is that we can
compute the Rao-Blackwellized estimator of joint and conditional
probabilities involving the central ranks without having to sum over
$p!^g$ elements. The key ingredient is the conditional independence of
the $\pi_i$'s given $\boldsymbol\theta.$ Thus for any subsets
$A_1,\ldots,A_g$ of $\mathfrak{S}_p,$ we can estimate the posterior
probability $P(\pi_i \in A_i\forall i ~|~ \mathbf{y})$ by
\[\frac{1}{M}\sum_{m=1}^{M} \prod_{i=1}^gP(\pi_i \in A_i | \tilde{\boldsymbol \theta}^{(m)},\mathbf{y}).\]
And although MCMC estimates of such statements could be found by
drawing samples from the conditional distribution of $\pi_i$'s given
theta, Rao-Blackwellization would result in estimates with smaller
MCMC variance.

Next, notice that there are two ways to compute conditional probabilities such as
\[P(\pi_i \in A_i,~\forall i ~|~ \pi_i \in B_i~\forall i,\mathbf{y})\]
where $A_i$'s and $B_i$'s are subsets of $\mathfrak{S}_p$. Either we
estimate it as the average of the conditional probabilities given
$\boldsymbol\theta$, i.e., by
\begin{align*}
&\frac{1}{M}\sum_{m=1}^{M}P(\pi_i \in A_i,~\forall i ~|~ \pi_i \in B_i~\forall i, ~\tilde{\boldsymbol\theta}^{(m)},\mathbf{y}) \\&= \frac{1}{M}\sum_{m=1}^{M}\dfrac{P(\pi_i \in A_i\cap B_i,~\forall i | \tilde{\boldsymbol\theta}^{(m)},\mathbf{y})}{P(\pi_i \in B_i,~\forall i | \tilde{\boldsymbol\theta}^{(m)},\mathbf{y})},
\end{align*}
or as the ratio of averages of joint probabilities each given
$\boldsymbol\theta$, i.e., by
\[\dfrac{\sum_{m=1}^{M}P(\pi_i \in A_i\cap B_i,~\forall i | \tilde{\boldsymbol\theta}^{(m)},\mathbf{y})}{\sum_{m=1}^{M}P(\pi_i \in B_i,~\forall i | \tilde{\boldsymbol\theta}^{(m)},\mathbf{y}) } \]
Although both are computationally feasible because given
$\boldsymbol\theta$ the $\pi_i$'s are independent we use the second
estimator in our data analysis in Section~\ref{sec:data}. The
second estimator has lower variance as there could be samples
$\boldsymbol\theta^{(m)}$ for which the probabilities $P(\pi_i \in
B_i,~\forall i | \boldsymbol\theta^{(m)})$ are infinitesimally small.

These joint and conditional posterior probabilities are important in
assessing how the preference for a particular item or a set of items
vary over different categories. Because the categories may be induced
by levels of factorial covariates these posterior probabilities
provide a way to compare the interactions between pairs of
factors. The simplest example would be to identify how the posterior
probability of an item being most preferred changes from one
cross-level interaction to another. We can then repeat the same
exercise on another item conditioning on the most preferred item and
develop more complicated probability statements that bring
out various aspects of a particular dataset.

We use a Monte Carlo EM algorithm based on the sandwich chain to
estimate the hyperparameter $\lambda$.
In the following, we
discuss this EM algorithm in detail.

\subsection{Estimating $\lambda$ via Monte Carlo
  EM}
  \label{sec:estimateLambda}

As mentioned before, we consider an
empirical Bayes approach for making inference of the
hyperparameter $\lambda$. In particular, we estimate $\lambda$ by
\[
\hat{\lambda} = \underset{\lambda}{\mbox{argmax}} \;c_\lambda(\mby),
\]
where
\[
c_\lambda(\mby) = \int \sum_{\boldsymbol\pi \in \mathfrak{S}_p^g} p(\mby | \boldsymbol\theta, \boldsymbol\pi) p(\boldsymbol\theta | \lambda) p(\boldsymbol\pi) d\boldsymbol\theta
\]
is the normalizing constant of the joint posterior density
$p(\boldsymbol\theta, \boldsymbol\pi | \mby)$ given in \eqref{eqn:posterior}.

Note that
\[
c_\lambda(\mby) =  \int \sum_{\boldsymbol\pi \in \mathfrak{S}_p^g}p(\mby, \boldsymbol\theta, \boldsymbol\pi | \lambda) d\boldsymbol\theta
\]
where $p(\mby, \boldsymbol\theta, \boldsymbol\pi | \lambda)$ is the joint probability
distribution of $\mby, \boldsymbol\theta$, and $\boldsymbol\pi$. This naturally leads
to an EM algorithm by treating $(\boldsymbol\theta, \boldsymbol\pi)$ as ``missing''
variables and considering a ``$Q$ function'' defined as
\[
Q(\lambda | \lambda') = \int\sum_{\boldsymbol\pi \in \mathfrak{S}_p^g}
\log p(\mby, \boldsymbol\theta, \boldsymbol\pi | \lambda) p (\boldsymbol\theta, \boldsymbol\pi | \mby,
\lambda') d\boldsymbol\theta .
\]
Then from Figure \ref{fig:graphical}, we see that
\[Q(\lambda|\lambda') = \int \log
p(\boldsymbol\theta| \lambda)~p(\boldsymbol\theta |\mby,
\lambda')d\boldsymbol\theta\] plus a term that does not depend on
$\lambda,$ so that we can avoid summing over
$\boldsymbol\pi\in\mathfrak{S}_p^g.$ Consequently, starting from an
initial estimate $\lambda^{(0)},$ the $(k+1)$st EM iterate of
$\lambda$ is given by maximizing $Q\left(\lambda|\lambda^{(k)}\right)$, that is
\begin{align}\label{eqn:Mstep}
  \lambda^{(k+1)} = \underset{\lambda}{\mbox{argmax}} \; \Bigg[ \sum_{i=1}^{p!}& e^{\lambda|\zeta_i|}~\mathrm{E}\left(\log\theta_i|\mby,\lambda^{(k)}\right)\\    - \sum_{i=1}^{p!} \log\Gamma\left(e^{\lambda|\zeta_i|}\right) &+ \log\Gamma\left(\sum_{i=1}^{p!} e^{\lambda|\zeta_i|}\right) \Bigg] \nonumber
\end{align}
Since the expectation term in \eqref{eqn:Mstep} is not available in
closed form, following \cite{case:2001}, we replace this expectation by
its estimate. To this end, suppose $\{\tilde{\boldsymbol\theta}^{(m)}:
m=1,\ldots,M\}$ is the sandwich chain from Section~\ref{sec:imda} having the
stationary distribution
$p\left(\boldsymbol\theta|\mby,\lambda^{(k)}\right).$
Then, \[\mathrm{E}\left(\log\theta_i|\mby,\lambda^{(k)}\right)
\approx (1/M)\sum_{m=1}^M\log\tilde{\theta}_{i}^{(m)}\] where $\tilde{\theta}_{i}^{(m)}$ is
the $i$th component of $\tilde{\boldsymbol\theta}^{(m)}.$
In our practical implementation, we run this stochastic version of the
EM with sandwich samples of $\boldsymbol\theta$ of moderate sizes until
the $\lambda^{(k)}$'s start fluctuating around the mode and then
perform a single EM iteration with a large sandwich chain of
$\boldsymbol\theta$ and obtain the final estimate $\hat\lambda.$
In order to compute the standard error  of $\hat\lambda,$ we use a
method described in \cite{case:2001}. The details of the standard error (se)
calculations are given in Appendix~\ref{app:lamse}.

\section{Simulation examples}
\label{sec:simu}
\begin{figure}[hp]
\begin{center}
\includegraphics[width=.8\linewidth]{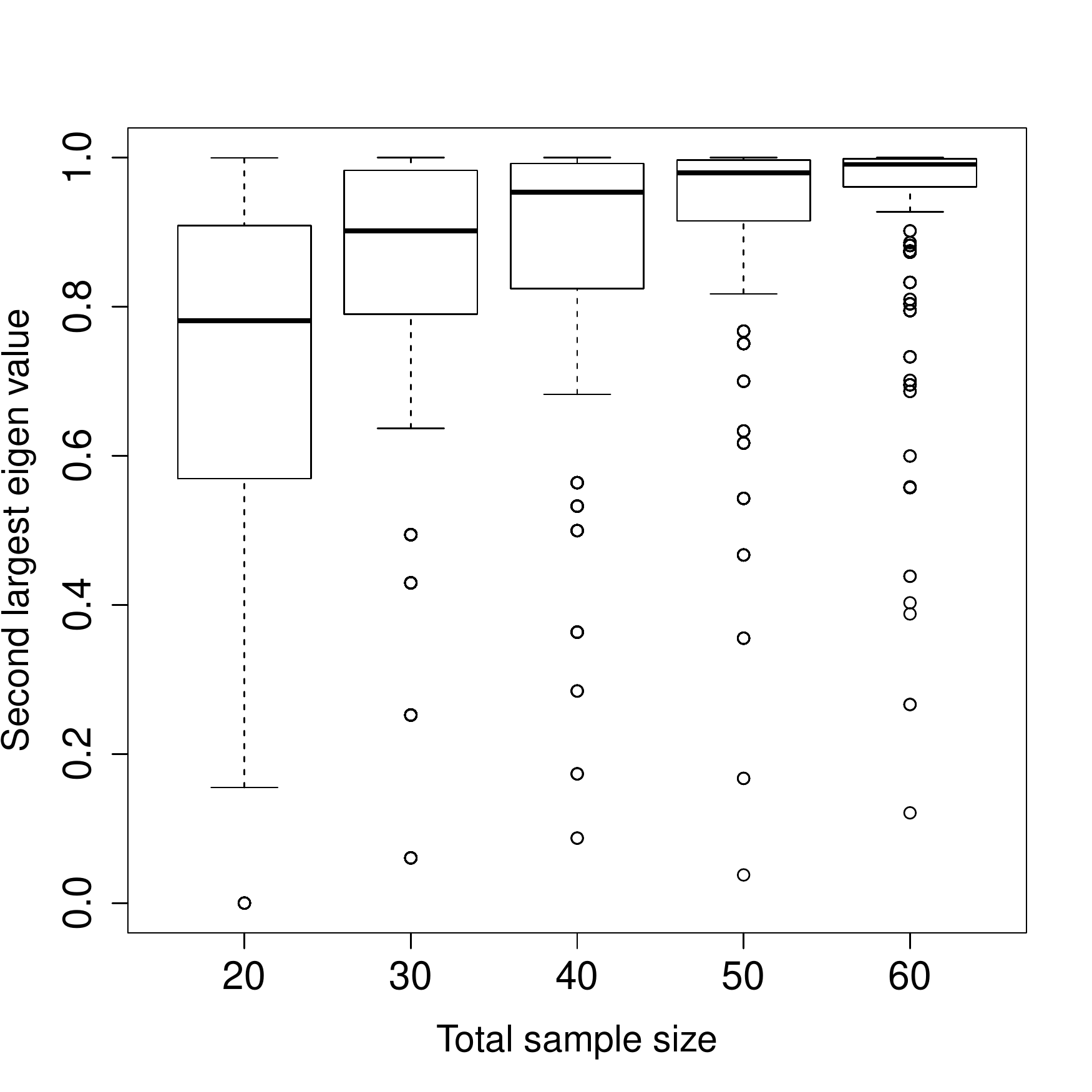}
\caption{The behavior of the second largest eigenvalue for the DA chain $\{\boldsymbol\pi^{(m)}\}_{m
  \ge 0}$.  The graph shows how the dominant eigenvalue of
  the DA chain changes with sample size, $n$.}
\label{fig:eig_gibbs}
\end{center}
\end{figure}

We study the performance of the Gibbs and sandwich chains through simulation
studies. Consider the situation where $p=2$ and $g=2$, that is, we
have two categories, two items to rank, and we let the number of
observations, $n$ vary. The small values of $p$ and $g$ allow us
to compute different Markov transition probabilities in closed form
and also it makes possible to compare the rank probability estimates with
their true values. Here we consider independent, uniform priors
on $(\pi_1, \pi_2)$, that is, $p(\pi_1, \pi_2) = p_1(\pi_1) p_2(\pi_2)$ with
\[
p_i(\zeta_1) = 0.5 = p_i(\zeta_2) \;\;\mbox{for}\; i=1,2.
\] We assume Beta $(a_1, a_2)$ prior on $\boldsymbol\theta$. We can
write down the Markov transition matrix (Mtm) $K_{\boldsymbol\pi}$ of the DA chain $\{\boldsymbol\pi^{(m)}\}_{m
  \ge 0}$ in closed form (see Appendix~\ref{app:mtm}). In order to
simulate data, we assume that the true ranks for category 1 and 2 are
$\zeta_1$ and $\zeta_2$ respectively. In this section, since we study
the convergence performance of the MCMC algorithms, we fix the value
of $\lambda$. In particular, we assume $\lambda= \log 2$, that is, we
have $a_1=2, a_2=1$. In our simulation study, we consider same number
of observations in each category. We calculate the entries of $K_{\boldsymbol\pi}$ by
numerical integration using the formula given in the
Appendix~\ref{app:mtm}.  For each fixed sample size we repeat the
simulation 1000 times, that is, we observe 1000 sets of observations
and calculate the corresponding Markov transition matrices. It is
known that the second largest eigenvalue ($\rho$) of $K_{\boldsymbol\pi}$ shows the
speed of convergence of the DA chain $\{\boldsymbol\pi^{(m)}\}_{m
  \ge 0}$ \cite[see e.g.][p. 209]{brem:1999} and hence of $\{\boldsymbol\theta^{(m)}\}_{m
  \ge 0}$ as well as of the Gibbs chain. Figure~\ref{fig:eig_gibbs} shows
the boxplots of one thousand $\rho$ values corresponding to each
sample size. The labels in the x-axis shows the number of observations
in each category. From the plot we see that the convergence rate of
the DA chain deteriorates as the sample size increases.

We now explain that the reason for the slow convergence of the DA chain is
its inability to move away from the local mode. We consider the case
where we have 50 observations in each category. In particular, we consider
one simulated data set where $n_{11} =40,  n_{12} = 14, n_{21} =10$, and $n_{22} =36$,
where $n_{ij}$ denote the number of observations in the
$j$th category with rank $\zeta_i$ for $i,j=1,2$.
The Mtm in this case is:

\[ K_{\boldsymbol\pi} = \left[ \begin{array}{cccc}
 0.0570291 & 0.7460761 & 0.1478273 & 0.0490667\\
 0.0000006 & 0.9999993 &
0.0000000 & 0.0000001\\ 0.0000004 & 0.0000001 & 0.9999981 & 0.0000014\\
 0.0574185 & 0.1962019 & 0.6818719 & 0.0645066\\
           \end{array} \right] \;.
\]

\begin{figure}[hp]
 \begin{center}
 \includegraphics[width=.9\linewidth]{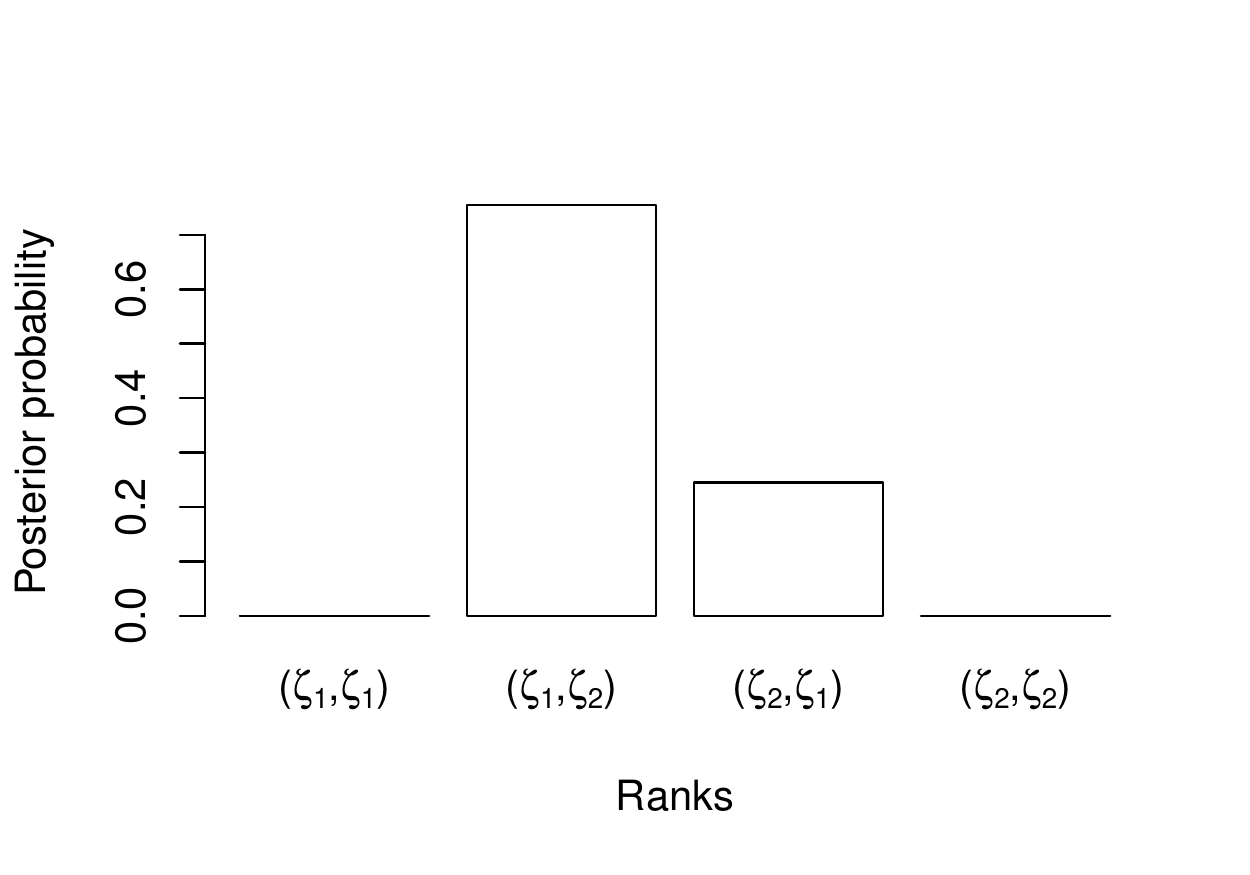}
 \caption{Plot of posterior density of $\boldsymbol\pi$.}
 \label{fig:jtpi}
 \end{center}
 \end{figure}

We have ordered the points in the state space as follows:
$(\zeta_1,\zeta_1)$, $(\zeta_1,\zeta_2)$, $(\zeta_2,\zeta_1)$, and
$(\zeta_2,\zeta_2)$. So, for example, the element in the second row,
third column is the probability of moving from $(\zeta_1,\zeta_2)$ to
$(\zeta_2,\zeta_1)$. Figure~\ref{fig:jtpi} shows the plot of the true
marginal posterior density of $\boldsymbol\pi$. From the plot we see that the
posterior density has its mode at $(\zeta_1,\zeta_2)$ with probability
around $0.75$. The rest of its probability mostly lies at
$(\zeta_2,\zeta_1)$ with almost no weight at the other two
ranks. Suppose we start the chain at $(\zeta_2, \zeta_1)$.  We expect
the chain to remain at $(\zeta_2, \zeta_1)$ for about $1/(1-0.9999981)
\approx 526,315$ iterations before it moves away to another state, that
is, the chain remains stuck in a low probability region (a local mode)
for a large number of iterations. Conditional on the chain leaving
$(\zeta_2, \zeta_1)$, the expected number of steps before it reaches
the true mode, $(\zeta_1, \zeta_2)$, is also quite large. Given that
the chain leaving $(\zeta_2, \zeta_1)$ there is about 74\% chance that
it moves to $(\zeta_2, \zeta_2)$ from where the probability that it
will move back to $(\zeta_2, \zeta_1)$ is about 0.68. Once the chain
is back at $(\zeta_2, \zeta_1)$, it is expected to stay there for
$526,315$ iterations before it jumps out to another state. All of this
translates into slow convergence.

\begin{figure}[htp]
\centering
\includegraphics[width=0.9\linewidth]{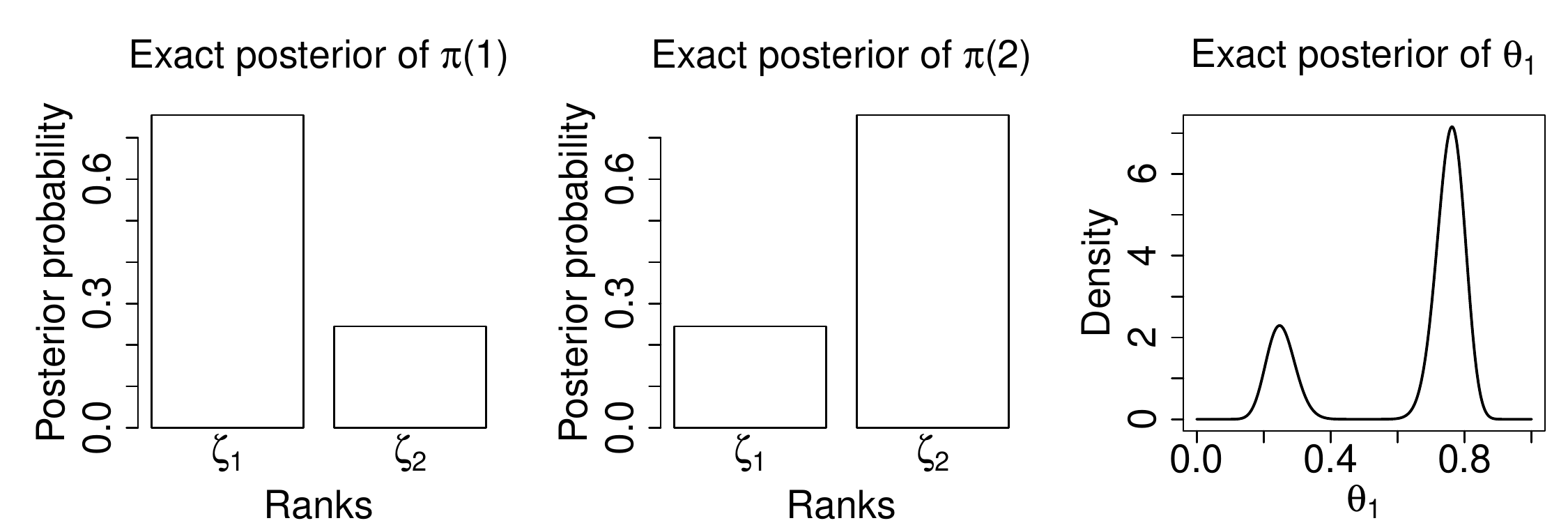}
\begin{center} (a) \end{center}
\includegraphics[width=0.9\linewidth]{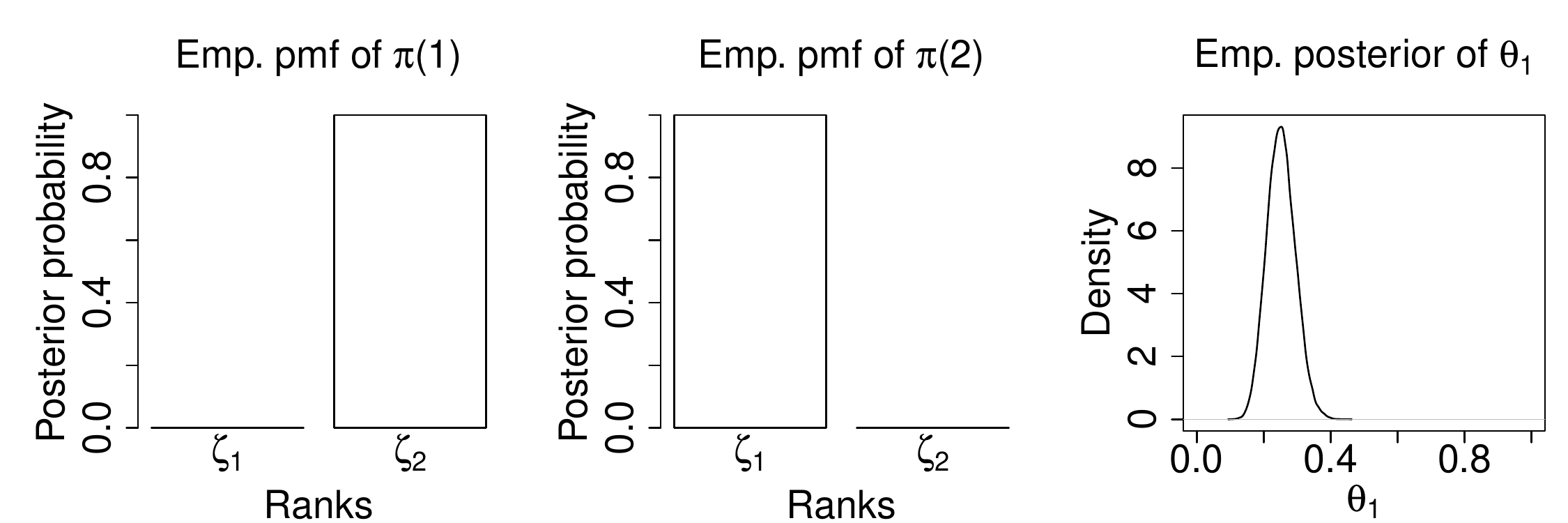}
\begin{center} (b) \end{center}
\includegraphics[width=.9\linewidth]{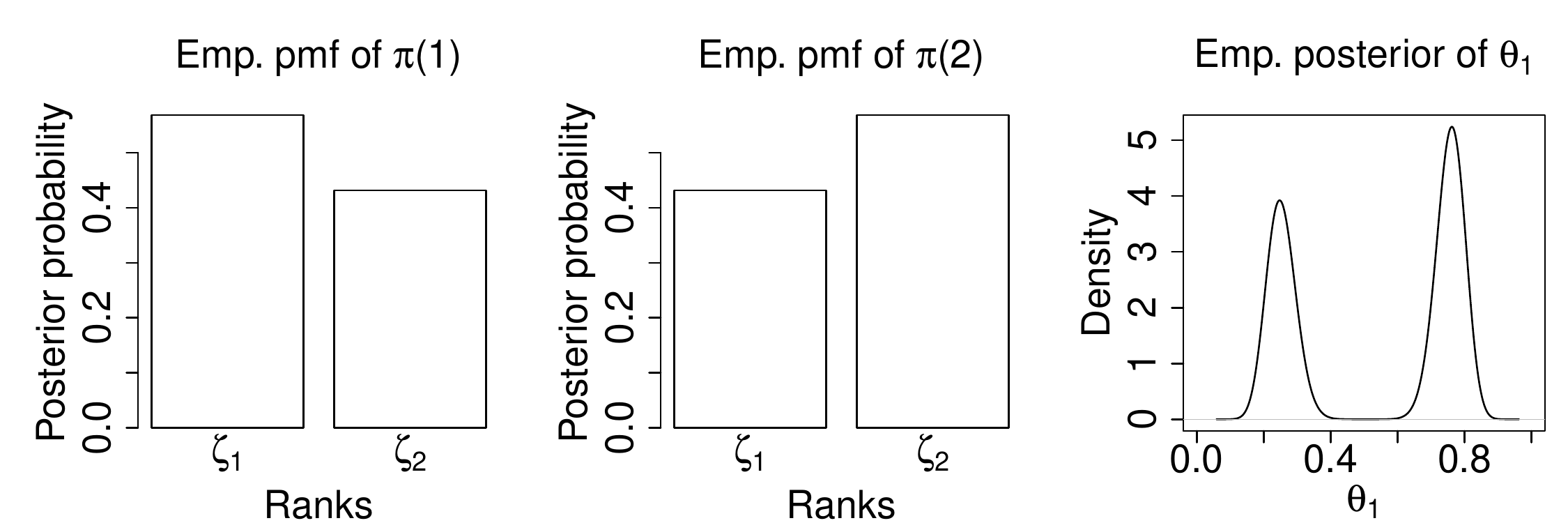}
\begin{center} (c) \end{center}
\includegraphics[width=.9\linewidth]{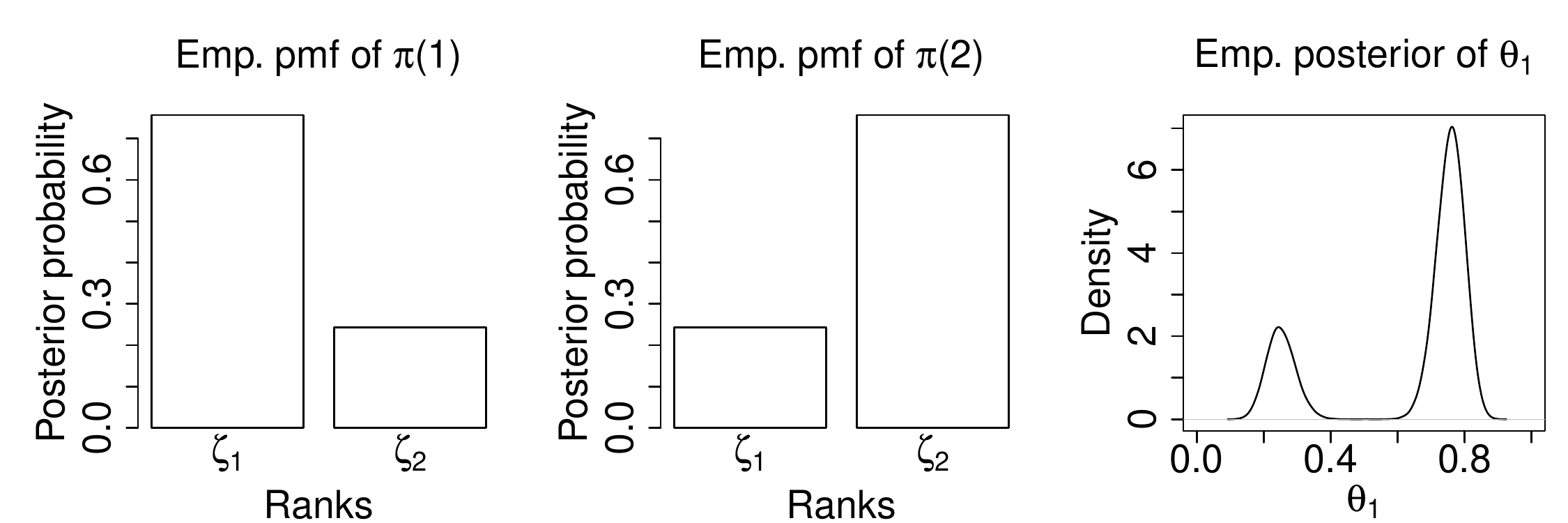}
\begin{center} (d) \end{center}
\caption{\small{(a) True posterior marginals of all parameters. Empirical estimates of the marginal densities based on the DA  chain with (b) 50 thousand iterations (c) 5 million iterations. (d) Empirical estimates of the marginal densities based on the sandwich chain with 50 thousand iterations.}}
\label{fig:simpost}
\end{figure}

 Next, we consider the same simulated data set to compare performance
 of the DA and sandwich chains. Since we are using small values of $p$
 and $g$, we can calculate the true posterior densities of
 $\boldsymbol\theta$ and $\boldsymbol\pi$ as given in
 Figure~\ref{fig:simpost} (a). From Figure~\ref{fig:simpost}, we see
 that although DA algorithm horribly fails to provide reasonable
 estimates even after 5 million iterations, the sandwich chain results
 in accurate estimates of true posterior probabilities in less than 50
 thousand iterations.  In fact, the Kolmogorov-Smirnov distance (plot
 is not included here) between the the true joint posterior
 distribution of $(\pi_1,\pi_2)$ and its empirical estimate based on
 the sandwich chain drops below $0.05$ within 100 iterations.

\begin{figure}[htp]
 \centering
\includegraphics[width=0.9\linewidth]{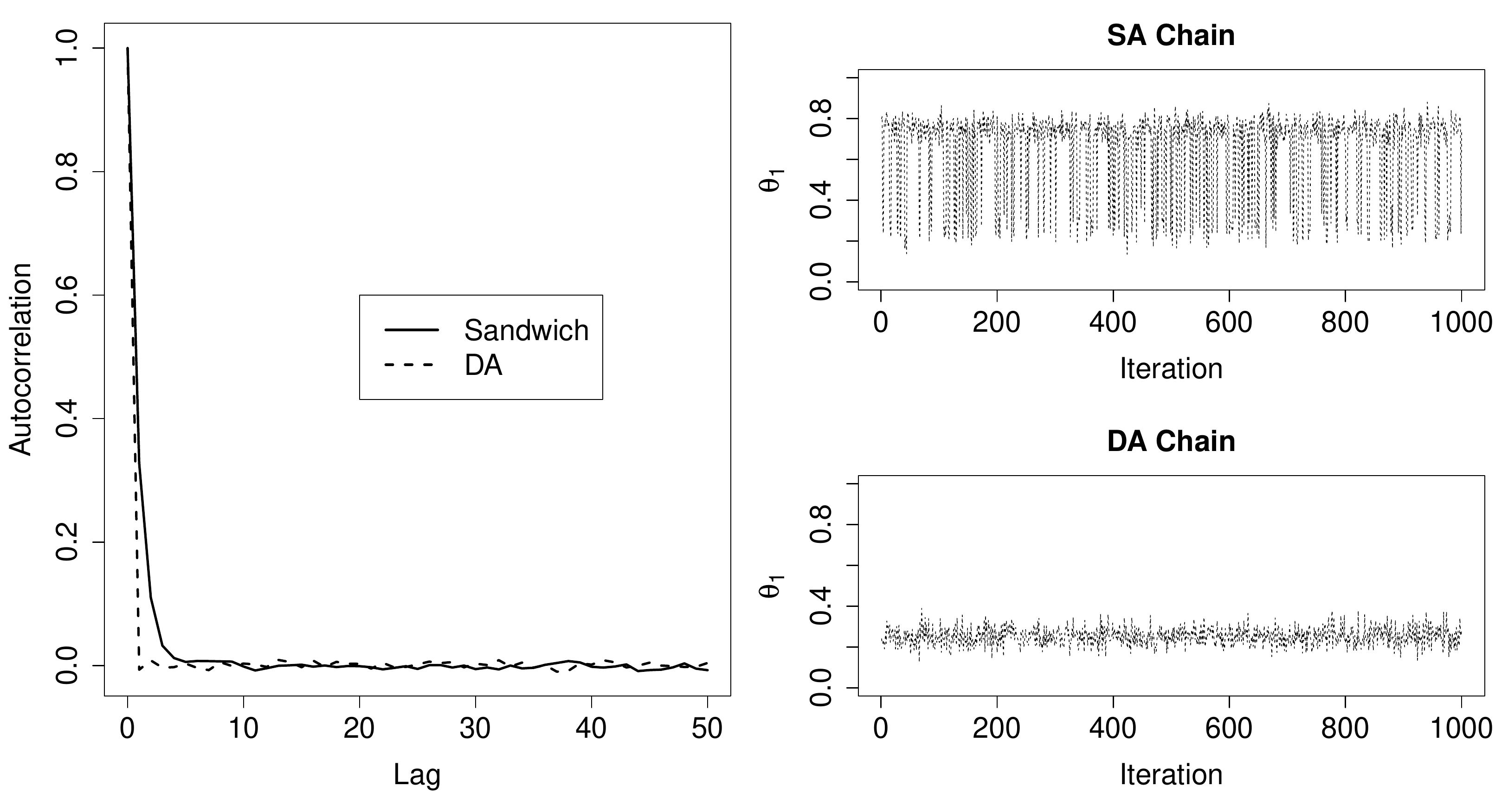}
\begin{center} (a) \end{center}
\includegraphics[width=0.9\linewidth]{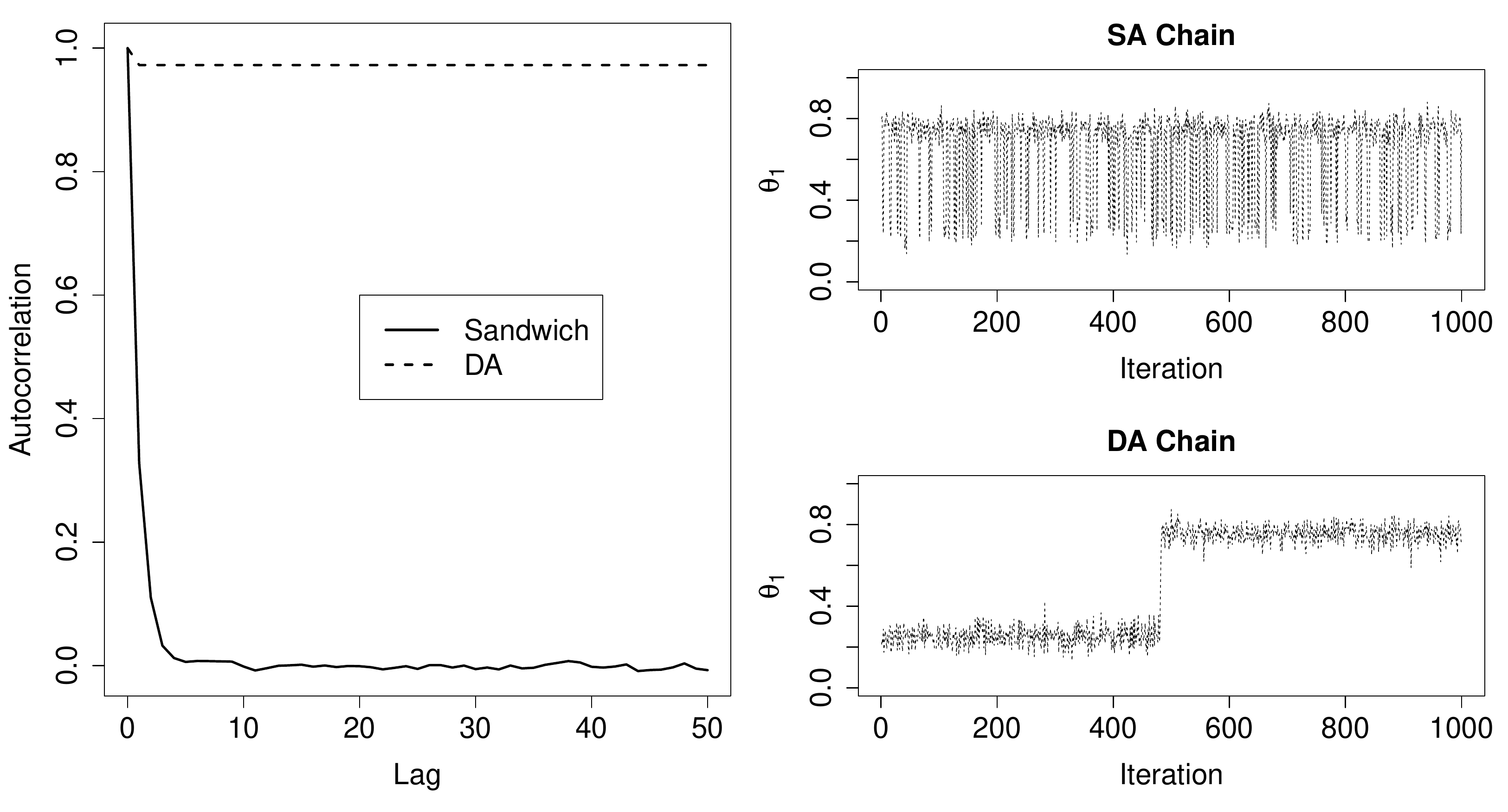}
\begin{center} (b) \end{center}
\caption{\small{Autocorrelation plots and trace plots for the DA and
sandwich $\boldsymbol\theta$ chains ($\theta_1$ component) based on (a) 50 thousand iterations
(trace plot shows last one thousand iterations) (b) 5 million iterations (trace plot shows 2,519,900 to 2,520,899 iterations).}}
\label{fig:acf}
\end{figure}

\begin{figure}[hbp]
\centering
\includegraphics[width=0.32\linewidth]{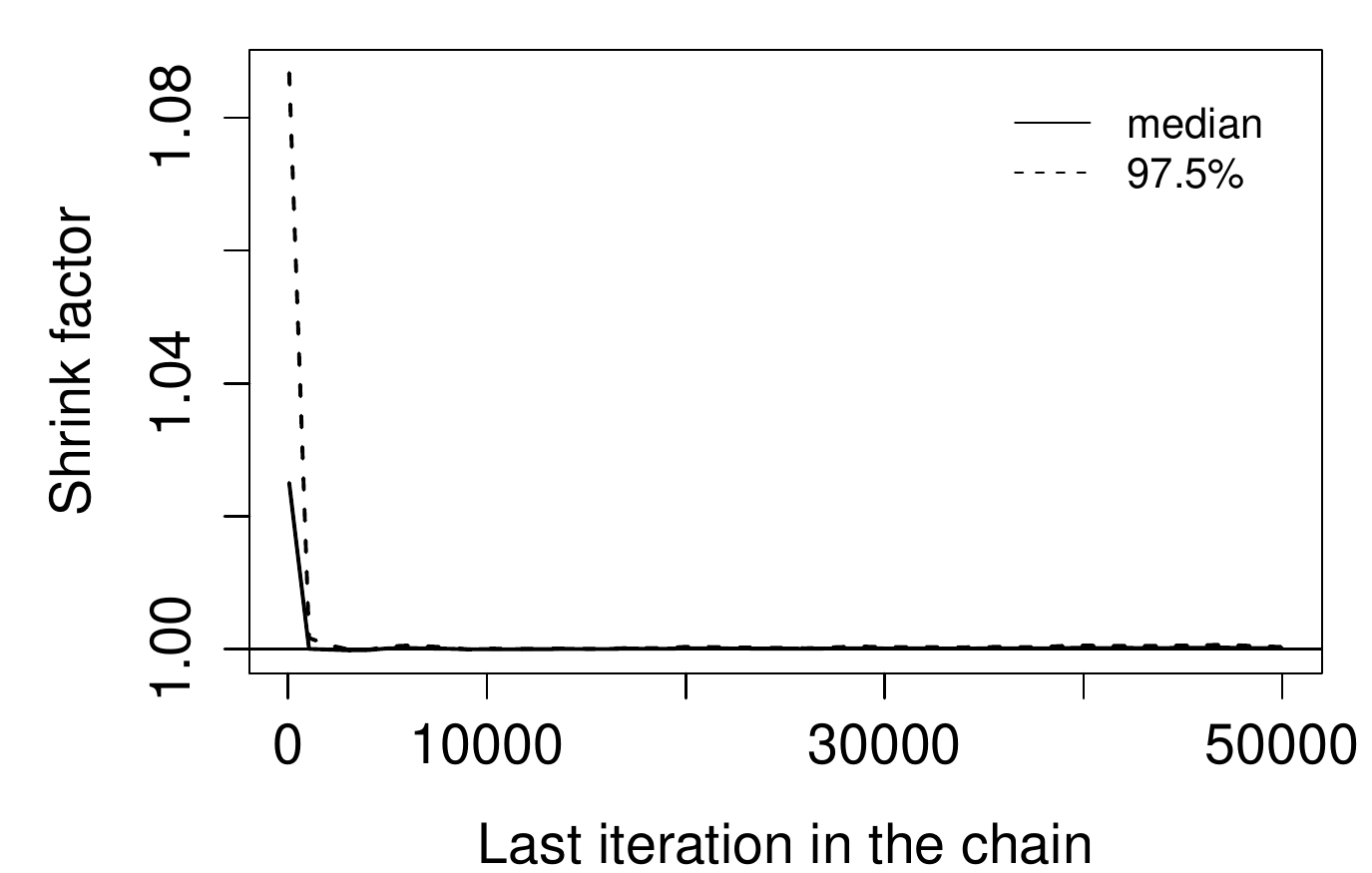}\includegraphics[width=0.33\linewidth]{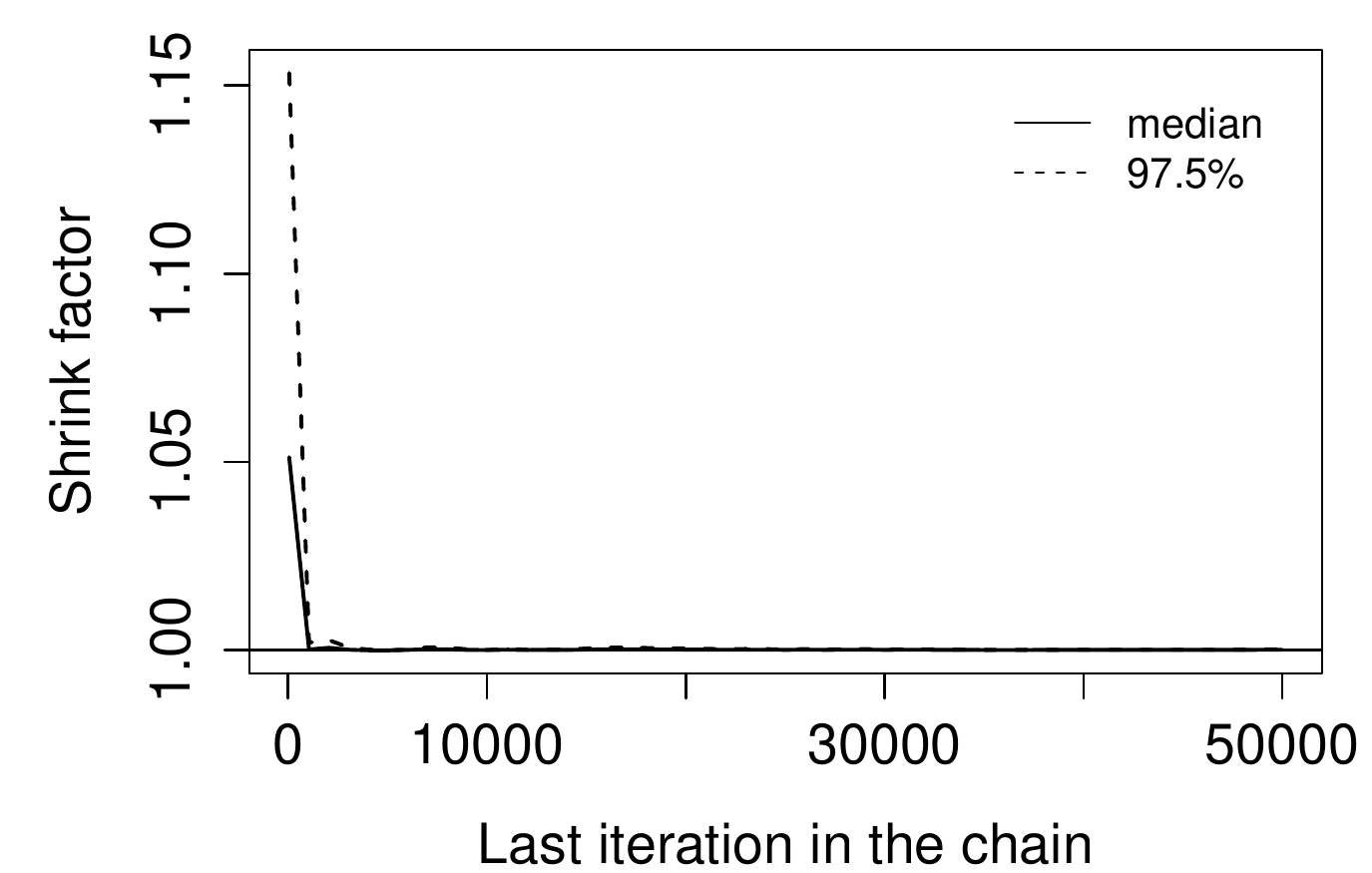}\includegraphics[width=0.33\linewidth]{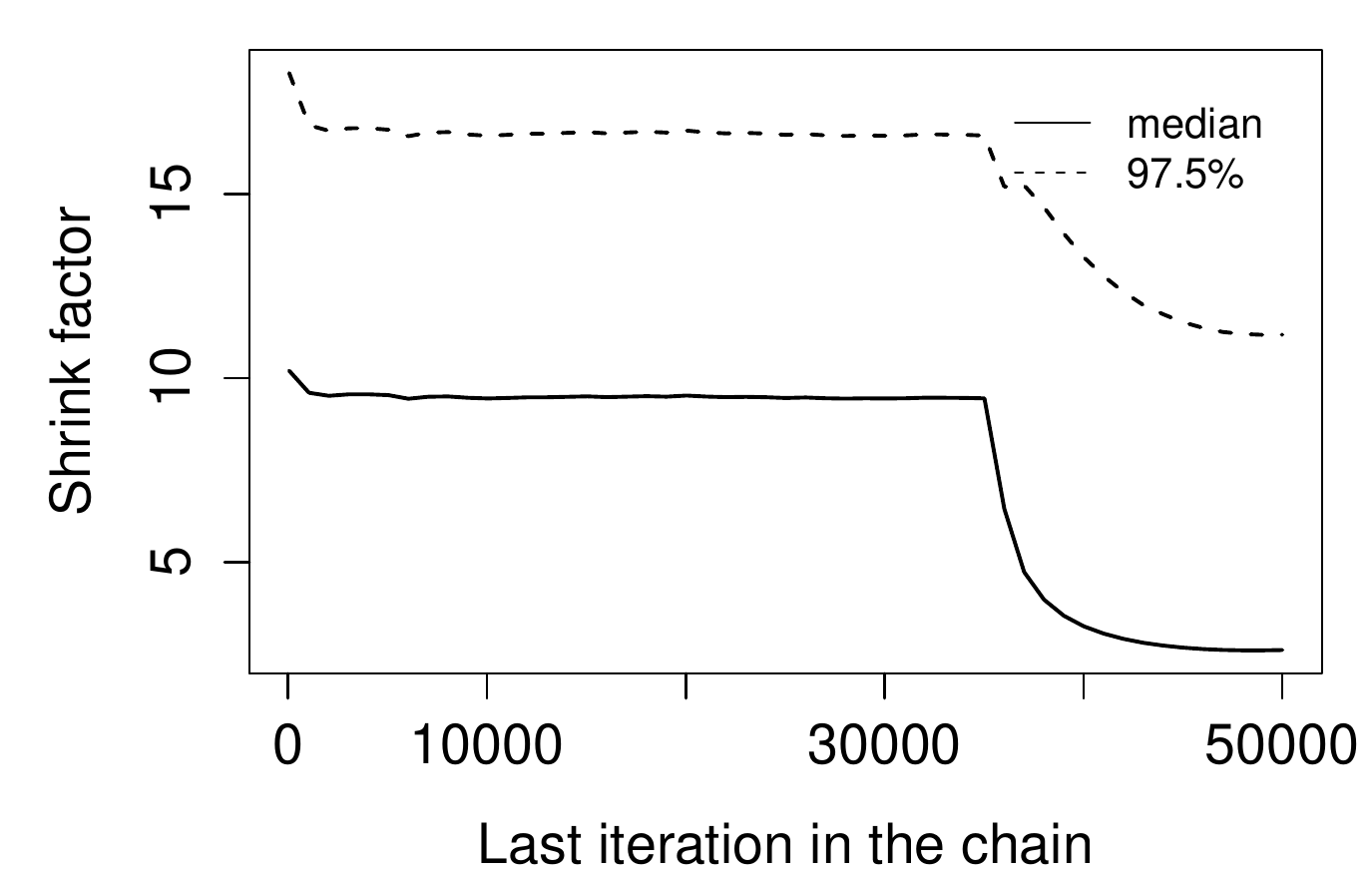}
\caption{\small{The Gelman-Rubin diagnostic plots for the DA chain based on 4 parallel runs of length 50,000 each. The starting values of multiple chains were selected
near the global mode (left) and the local mode (middle) and from both
modes.
}}
\label{fig:gelmanplot}
\end{figure}

 We now show how the empirical convergence diagnostics like, the trace
 plots and the autocorrelation plots can mislead MCMC practitioners by
 giving false impression that the chain has converged while the chain
 has failed to visit the mode of the posterior density even
 once. Figure~\ref{fig:acf} shows the autocorrelation plots and trace
 plots for the DA and sandwich chains based on 50 thousand and 5
 million iterations. From the trace plot in Figure~\ref{fig:acf} (a),
 it may seem that the DA chain is mixing well, which is corroborated
 by the corresponding autocorrelation plot. In fact, the
 autocorrelations for the DA chain is close to zero in less than three
 iterations, and they die down faster than the autocorrelations for
 the sandwich chain. Note that the Gibbs sampler has not been able to
 move between the local modes in 50,000 iterations and from
 Figure~\ref{fig:simpost} we know that the empirical estimates of
 probability mass functions (pmfs) and densities obtained using DA
 chains are far from true posterior probabilities. Since in practice,
 the true target densities are not available, MCMC practitioners may
 be misled by the empirical convergence diagnostics like trace plots
 and autocorrelation plots. When we run the Gibbs chain much longer (5
 million iterations), it eventually visits the other local mode. The
 right panel of Figure~\ref{fig:acf} (b) shows the trace plots for the
 DA and sandwich chains between 2,519,900 and 2,520,899 iterations (to
 capture a jump of the DA chain from one mode to other).  The left
 panel of Figure~\ref{fig:acf} (b) now shows the huge gains in
 autocorrelation by running the sandwich chain over the DA
 algorithm. It shows that even 50-lag autocorrelation for the DA chain
 is close to 1. Finally, we consider the popular potential scale
 reduction factor (PSRF) (\cite{gelm:rubi:1992}) for monitoring
 convergence of the above DA chain.  In general, calculation of PSRF
 begins with running multiple MCMC chains started at different
 (overdispersed) initial points. At convergence, these chains produce
 samples from the same distribution, and this is assessed by comparing
 the means and variances of these individuals chains with that of the
 pooled chain. If the PSRF is close to one, it is used as an indicator
 of the convergence to the stationarity. Figure~\ref{fig:gelmanplot}
 shows PSRF based on four parallel DA ($\boldsymbol\theta$) chains for
 50 thousand iterations with different starting values. From
 Figure~\ref{fig:gelmanplot} we see that when all chains are started
 close to one mode then PSRF diagnostics fails as the chains have not
 traveled the whole space yet. PSRF is able to indicate
 non-convergence of DA only when chains are started at different
 modes. Since, in practice, especially in multivariate settings, one
 does not have knowledge about the locations of the modes, PSRF may
 fail to catch convergence problems of MCMC algorithms.

\section{Real data application}
\label{sec:data}
In this section we illustrate the methods proposed in this article by
applying them to a sushi preference data, previously published in the
literature. The sushi preference data collected by \cite{kami:2003}
consists of complete rankings of few types of sushis by 5,000
respondents together with demographic data on the respondents. This
dataset has also been used by \cite{kami:akah:2009} for model based
clustering of orderings. In this article, we consider four types of
fish sushis: Anago (sea eel), Maguro (tuna), Toro (fatty tuna) and
Tekka Maki (tuna roll). We study the role of gender (male or female),
age and geographic location (east or west Japan) of the respondents on
ranking of these four types of sushis. The variable age is categorized
into six categories: 15--19 years, 20--29 years, 30--39 years, 40--49
years, 50--59 years and 60 years or above. Thus the respondents are
categorized into overall 24 categories. Although Toro (fatty tuna)
stands out as the most preferred sushi in all categories, substantial
variability is observed in ranking the remaining three sushis. Thus
apart from identifying the central ranking for these categories we are
also interested in the heterogeneity in ranking of Anago, Maguro and
Tekka Maki conditional on the Toro being the most preferred sushi. The
MCMC sampling outputs allow us to compute these conditional
probabilities without having to sum over an enormous $24^{24}$
dimensional joint distribution.

To this end, we ran our sandwich algorithm for a total of 60,000
iterations and discarded the first 10,000 iterations as burnins,
although convergence was observed within first 15,000 iterations. As
far as the marginal probabilities are concerned, the central ranks are
same for all the categories and it is
\begin{center}
 Toro $\succ$ Maguro $\succ$ Tekka Maki $\succ$ Anago
\end{center}
where Toro $\succ$ Maguro means Toro is preferred to Maguro. However,
there is substantial heterogeneity in ranking. The marginal
probability that Toro is the most preferred sushi ranges from 66\% to
69\% except for females of age 60 years or more and currently living
in west Japan, where the probability is 44.3\%. However, the sample
sizes are relatively small in these two categories -- there are only
12 females in the east Japan of age 60 or above and only 5 females in
the west of age 60 or above. Thus the marginal posterior distributions
of the central ranks for these two categories have higher variability
than the other groups. The joint (posterior) probability of Toro being
the most favorite sushi across all the categories is 37\% (se 0.2\%). We calculate the Monte Carlo standard errors for the posterior
estimates using batch means method \citep{fleg:jone:2010}. The joint
(posterior) probability of Toro being among the top two favorite
sushis across all the categories is 78\% (se 0.3\%). Thus Toro
can be regarded as the most preferred sushi among the four.

Next assuming Toro being the most preferred, we compute the
conditional probabilities of other sushis being the second most
favorite in each of the groups. These conditional probabilities are
highest for Maguro (tuna) and varies from 49\% to 52\% among the
categories, followed by Anago (sea eel) with probabilities varying
from 24\% to 35\%. The probability of Maguro being the second most
favorite across all categories given that the Toro is the most
favorite across all categories is 47\% while the same probability for
Anago is only 26\%. 

\begin{figure}[hp]
 \centering\includegraphics[width=0.5\textwidth]{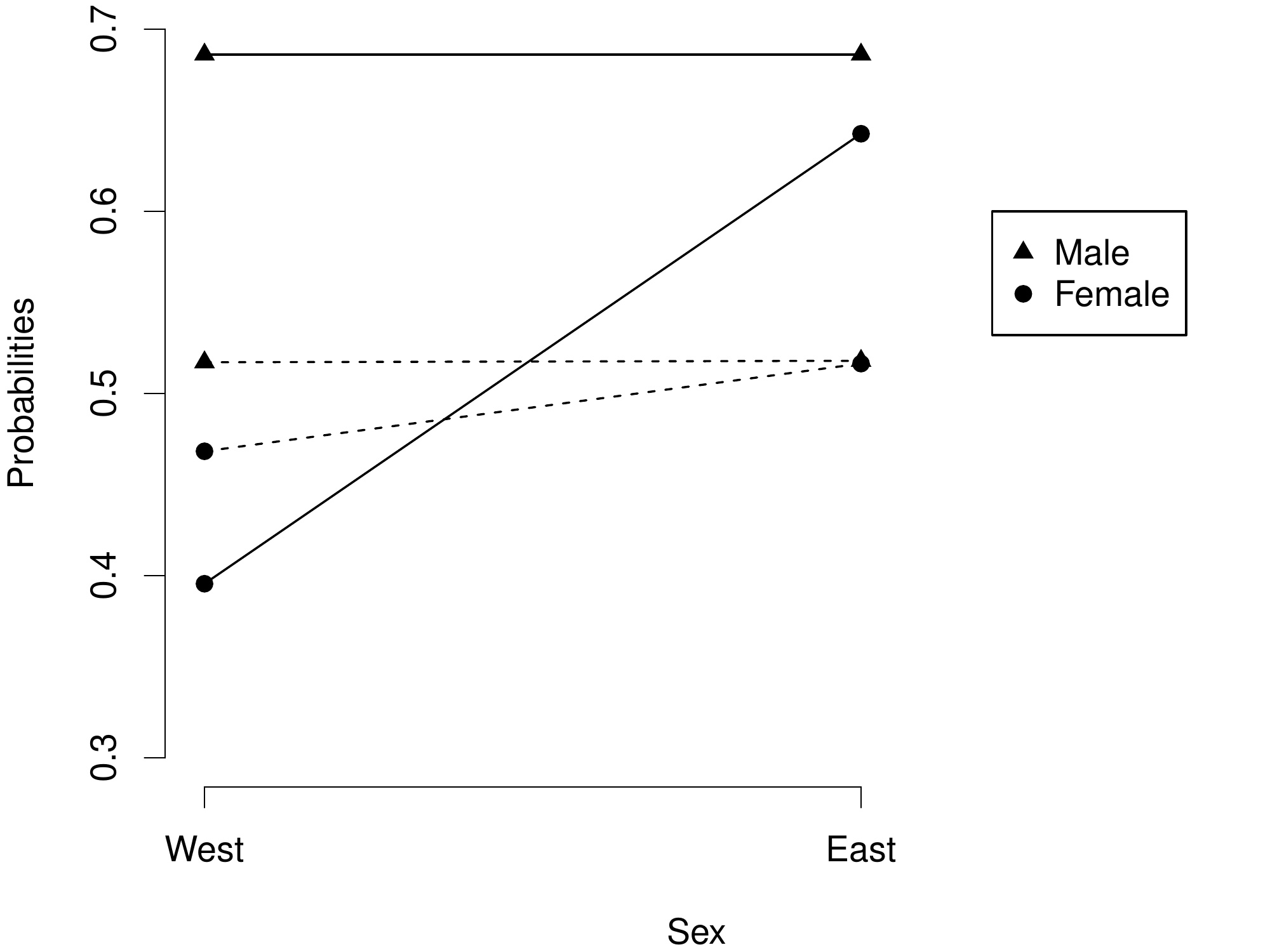}
 \caption\small{{Interaction probabilities: the solid lines join the points
   denoting the posterior probabilities of Toro being the most
   favorite sushi across all age groups. The dashed lines join the
   points denoting the posterior probabilities of Maguro being the
   second most favorite sushi conditional on Toro being the most
   favorite across all age groups.}}
 \label{fig:interaction}
\end{figure}

In order to judge any interaction effect present we compute the
probability that the Toro is the most favorite sushi across all the
age groups for each combination of gender (male/female) and geographic
location (east/west). The solid lines in Figure \ref{fig:interaction}
display these posterior probabilities for males (triangles) and
females (circles). Notice that the posterior probability of Toro being
the most favorite sushi among men does not change from east to west
Japan but for women it does. Similarly we compute the posterior
interaction probabilities for Maguro being the second most favorite
conditional on Toro being the most favorite. The dashed lines in
Figure \ref{fig:interaction} display these probabilities. Similar to
the previous case, the conditional posterior probability changes for
women from east to west Japan. However, this change is not as large as
the previous case. Such interaction effects were found to be absent
between age and gender, suggesting that the preferences towards
different sushis do not, as such, depend on age.

\begin{figure*}[htp]
 \centering\includegraphics[width=0.8\textwidth]{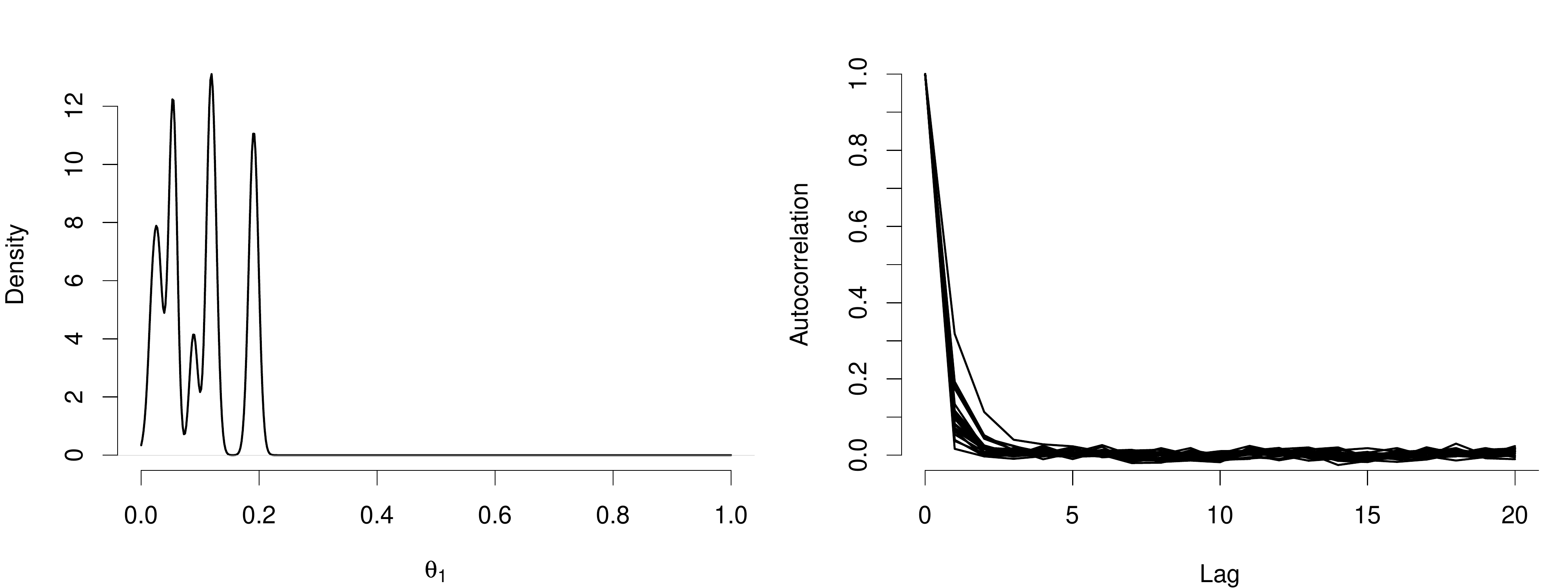}
 \caption{\small{Left: Multimodal posterior density of $\theta_1$. Right: Autocorrelation plots of the $\theta_i$ chains after thinning by 5.}}
 \label{fig:sushiPlots}
\end{figure*}

The variability in ranking is also reaffirmed by the small value of
$\widehat\lambda = 0.175$ (se 0.235) and the multi-modal posterior
distribution of $\theta_i$'s. For example, the posterior density
estimate of $\theta_1$ is shown in the left panel of Figure
\ref{fig:sushiPlots}. It is important to note that exploring this
variability in ranking is possible because our sandwich algorithm for
$\boldsymbol\theta$ mixes very well (see right panel of Figure
\ref{fig:sushiPlots}) and discovers all the modes of the posterior
distributions of $\theta_i$'s.

\section{Discussions}
\label{sec:disc}
In comparison to the multinomial logit model, the new rank-data model
presented in this paper is far more parsimonious and it requires
estimation of far less number of parameters. In addition, the new
rank-data model allows us to introduce a notion of central ranking
that facilitates a natural modeling of the observed rank data as
perturbations of the central rank. In the presence of categorical
covariates the central rank may vary across categories and the
rank-data model has been extended to incorporate this. While the
estimation of the parameters of this new rank-data model is
conceptually straight forward in a Bayesian set-up, considerable
computational challenges are encountered due to the very slow
convergence of the Gibbs sampler. Widely used
diagnostics for detecting the convergence of the Gibbs sampler
seem to fail in this situation. A new sandwich algorithm is
devised for efficient posterior computation and
is seen to perform effectively in the practically important situation
when the number of items to be ranked is small.  The case when one or
more covariates are continuous will require development of a different
technique and will be taken up in a future paper. We are also going to apply the proposed model and the MCMC
algorithms in the context of partial rank data, that is, when not all
available items are ranked.

\section{Supplementary materials}
The online supplementary materials contain codes for analyzing the
sushi data.

\medskip

\noindent {\Large \bf Appendices}

\begin{appendix}
\section{Standard errors of $\hat\lambda$}
\label{app:lamse}
To obtain the standard error of
$\hat\lambda,$ note that the negative of the observed information for
$\lambda$ is given by
\begin{align*}
 & -I(\lambda|\mby) = \frac{d^2}{d\lambda^2}\log c_\lambda(\mby)\\ & =  \mathrm{E}~\left(\frac{d^2}{d\lambda^2}\log p(\mby,\boldsymbol\theta,\boldsymbol\pi|\lambda) \bigg|~ \lambda \right) + \mathrm{Var}\left(\frac{d}{d\lambda}\log p(\mby,\boldsymbol\theta,\boldsymbol\pi|\lambda) \bigg|~\lambda\right)  \\
  & =  \mathrm{E}~\left(\frac{d^2}{d\lambda^2}\log p(\boldsymbol\theta|\lambda) \bigg|~ \lambda \right) + \mathrm{Var}\left(\frac{d}{d\lambda}\log p(\boldsymbol\theta|\lambda) \bigg|~\lambda\right) \\
  & =  \sum_{i=1}^{p!}|\zeta_i|^2 e^{\lambda|\zeta_i|} \left\{\mathrm{E}(\log\theta_i|\mby,\lambda) - \Psi_{1}\left(e^{\lambda|\zeta_i|}\right)e^{\lambda|\zeta_i|} - \Psi\left(e^{\lambda|\zeta_i|}\right)\right\}  \\
  & \qquad + \Psi_1\left(\sum_{i=1}^{p!}e^{\lambda|\zeta_i|}\right)\left(\sum_{i=1}^{p!}e^{\lambda|\zeta_i|}\right)^2 + \Psi\left(\sum_{i=1}^{p!}e^{\lambda|\zeta_i|}\right)\sum_{i=1}^{p!}e^{\lambda|\zeta_i|} \\
  &  \qquad+ \mathrm{Var}\left(\sum_{i=1}^{p!}|\zeta_i|e^{\lambda|\zeta_i|}\log\theta_i ~\bigg|~\mby,\lambda \right),
\end{align*}
where the first equality follows from \cite[][p. 497]{case:2001}, the
second equality follows from Figure~\ref{fig:graphical}, and $\Psi(\cdot)$
and $\Psi_1(\cdot)$ are the digamma and the trigamma functions
respectively. In the above, we approximate the conditional expectation
and variance terms by the corresponding MCMC estimate using our
sandwich chain. Consequently, the standard error of $\hat\lambda$ is
given by s.e.$(\hat\lambda) = I(\hat\lambda|\mby)^{-1/2}$.

\section{The Mtm $K_{\boldsymbol\pi}$ when $p=2, g=2$ }
\label{app:mtm}

In order to write down the Mtm $K_{\boldsymbol\pi}$ of the DA chain
$\{\boldsymbol\pi^{(m)}\}_{m \ge 0}$, we need to introduce some
notations.  Let $n_{ij}$ denote the number of observations in the
$j$th category with rank $\zeta_i$ for $i,j=1,2$. Let $n_{i.}= n_{i1}
+ n_{i2}$ for $i=1,2$, $n_{d}= n_{11} + n_{22}$, and $n_{od}= n_{12}
+ n_{21}$. Let $r(x) =
1/[x^{n_{1.}}(1-x)^{n_{2.}}+x^{n_{2.}}(1-x)^{n_{1.}}+x^{n_{d}}(1-x)^{n_{od}}+x^{n_{od}}(1-x)^{n_{d}}]$,
$a=1/\mbox{Beta} (n_{1.} + a_1, n_{2.} + a_2)$, $b=1/\mbox{Beta}
(n_{d} + a_1, n_{od} + a_2)$, $c=1/\mbox{Beta} (n_{od} + a_1, n_{d} +
a_2)$, and $d=1/\mbox{Beta} (n_{2.} + a_1, n_{1.} + a_2)$.  Let
$k_{ij}$ be the $(i,j)$th element of the matrix $K_{\boldsymbol\pi}$, $i,j=1,2,3,4$.
Then straightforward calculations show that
\[
k_{11} = a\int_0^1 r(x) x^{2 n_{1.} +a_1
-1} (1-x)^{2 n_{2.} +a_2 -1} dx\]\[k_{12} =a\int_0^1 r(x) x^{n_{1.} + n_d +a_1
-1} (1-x)^{n_{2.} + n_{od} + a_2 -1} dx
\]
\[
k_{13} = a\int_0^1 r(x) x^{n_{1.} + n_{od} +a_1
-1} (1-x)^{n_{2.} + n_{d} + a_2 -1} dx\]\[ k_{14} = a\int_0^1 r(x) x^{n +a_1
-1} (1-x)^{n +a_2 -1} dx
\]
\[
k_{21} = b\int_0^1 r(x) x^{n_{1.} + n_d +a_1
-1} (1-x)^{n_{2.} + n_{od} +a_2 -1} dx \]\[k_{22} =b\int_0^1 r(x) x^{2 n_d +a_1
-1} (1-x)^{2 n_{od} + a_2 -1} dx
\]
\[
k_{23} = b\int_0^1 r(x) x^{n +a_1
-1} (1-x)^{n + a_2 -1} dx\]\[ k_{24} = b\int_0^1 r(x) x^{n_{2.} + n_d +a_1
-1} (1-x)^{n_{1.} + n_{od} +a_2 -1} dx
\]
\[
k_{31} = c\int_0^1 r(x) x^{n_{1.} + n_{od} +a_1
-1} (1-x)^{n_{2.} + n_{d} +a_2 -1} dx\]\[k_{32} =c\int_0^1 r(x) x^{n +a_1
-1} (1-x)^{n + a_2 -1} dx
\]
\[
k_{33} = c\int_0^1 r(x) x^{2 n_{od} +a_1
-1} (1-x)^{2 n_{d} + a_2 -1} dx\]\[ k_{34} = c\int_0^1 r(x) x^{n_{2.} + n_{od} +a_1
-1} (1-x)^{n_{1.} + n_{d} +a_2 -1} dx
\]
and finally
\[
k_{41} = d\int_0^1 r(x) x^{n +a_1
-1} (1-x)^{n +a_2 -1} dx \]\[k_{42} =d\int_0^1 r(x) x^{n_d + n_{2.} +a_1
-1} (1-x)^{n_{od} + n_{1.} + a_2 -1} dx
\]
\[
k_{43} = d\int_0^1 r(x) x^{n_{2.} + n_{od} +a_1
-1} (1-x)^{n_{1.} + n_{d} + a_2 -1} dx\] \[ k_{44} = d\int_0^1 r(x) x^{2 n_{2.} +a_1
-1} (1-x)^{2 n_{1.} +a_2 -1} dx .
\]
As mentioned before, here we have ordered the points in the state space as follows:
$(\zeta_1,\zeta_1)$, $(\zeta_1,\zeta_2)$, $(\zeta_2,\zeta_1)$, and
$(\zeta_2,\zeta_2)$. So, for example, the element $k_{23}$ is the
probability of moving from $(\zeta_1,\zeta_2)$ to $(\zeta_2,\zeta_1)$.
Note that all of the transition probabilities are strictly positive,
which implies that the DA chain is Harris ergodic.
\end{appendix}

\section*{Acknowledgment}
The second author thanks the Indian Institute of Management,
Ahmedabad, India for kind support during his stay. The authors thank an editor, an associate editor
and a referee for their comments which improved the paper.

\end{document}